\documentclass[fleqn]{article}% fleqn ensures left-alignment of equations
\usepackage[top=1in,bottom=1.2in,left=1in,right=1in]{geometry} % 
\usepackage{mathtools}% loads amsmath
\usepackage{braket}
\usepackage{amssymb}% e.g. \gtrsim
\usepackage{array}
\usepackage{booktabs}
\usepackage{mathrsfs}% e.g. nice Lagrange-L with \mathscr{L}
\usepackage{dsfont}% BlackBoard letters
\usepackage{cite}%
\usepackage{geometry}%
\usepackage{tikzpagenodes}% pictures, loads xcolor, graphicx
\usetikzlibrary{shadows.blur}
\usepackage{wrapfig}%
\usepackage{nicefrac}% "nice" fraction
\usepackage[hidelinks]{hyperref}% hidelinks avoids the colored boxes
\usepackage[noabbrev,capitalise]{cleveref}% wants to be loaded last, use \Cref
\usepackage{float}
\usepackage[nolist]{acronym}
\usepackage{xspace}
\usepackage[small]{caption}
\usepackage{subcaption}%
\usepackage{mleftright}
\mleftright%
\makeatletter
\g@addto@macro\bfseries{\boldmath}
\makeatother
\newcommand{\rep}[1]{\ensuremath{\boldsymbol{#1}}}

\DeclareMathOperator{\re}{Re}
\DeclareMathOperator{\im}{Im}

\newcommand{\ii}{\hskip0.1ex\mathrm{i}\hskip0.1ex}% imaginary i, ISO conventions

\newcommand{\SO}[1]{\ensuremath{\mathrm{SO}(#1)}}
\newcommand{\SU}[1]{\ensuremath{\mathrm{SU}(#1)}}

\newcommand{\U}[1]{\ensuremath{\mathrm{U}(#1)}}
\newcommand{\Z}[1]{\ensuremath{\mathds{Z}_{#1}}} % z_N ->\Z{N}

\newcommand{\Hu}{\ensuremath{H_u}}
\newcommand{\Hd}{\ensuremath{H_d}}
\newcommand\AfourFlavonA{\Phi_\nu}
\newcommand\AfourFlavonB{\Phi_e}
\newcommand{\CP}{\ensuremath{\mathcal{CP}}\xspace}
\newcommand{\ChargeC}{\ensuremath{\mathcal{C}}}
\def\mytitle{Neutrino Flavor Model Building and the Origins of Flavor and \CP Violation: A Snowmass White Paper}
\numberwithin{equation}{section}
\numberwithin{figure}{section}
\numberwithin{table}{section}
\newcommand\snowmass{\vspace*{-1cm}%
\begin{center}
  \rule[-0.2in]{\hsize}{0.01in}\\
  \rule{\hsize}{0.01in}\\
  Submitted to the Proceedings of the US Community Study\\ 
  on the Future of Particle Physics (Snowmass 2021)\\
    \vskip 0.05in
    {\itshape Snowmass~2021 Theory Frontier \& Neutrino Frontier}\\
  \rule{\hsize}{0.01in}\\
  \rule[+0.2in]{\hsize}{0.01in}\\[-2em]
\end{center}
}
\usepackage[firstpage=true]{background}
\backgroundsetup{contents={\parbox{6.5in}{\snowmass}}, scale=1,
placement=top,opacity=1,color=black,position={3.25in,0.5in}}

\usepackage{fancyhdr}
\fancypagestyle{plain}{%
  \fancyhf{}%
  \fancyhead[C]{}
  \fancyfoot[C]{\thepage}
}

\fancypagestyle{title}{%
  \fancyhf{}%
  \fancyhead[C]{\itshape Submitted to the Proceedings of the US Community
  Study\\ on the Future of Particle Physics (Snowmass 2021)}
  \fancyfoot[C]{\thepage}
}
\begin{document}
\begin{titlepage}
\vspace*{0.2cm}

\begin{flushright}
UCI--TR--2022--03%
\\
Submitted to Snowmass TF11: Theory of Neutrino Physics; NF03: Beyond the Standard Model
\end{flushright}

\vspace*{0.1cm}
\linespread{1.15}
\begin{center}
\par{\Large\sffamily\bfseries\mytitle\par}%
\end{center}
\linespread{1}
\begin{center}
\vspace*{0.5em}
\newcounter{authorno}%
\setcounter{authorno}{0}%
\newcommand*{\authorpic}{\ifcase\value{authorno}\relax
\or
\includegraphics[height=2ex]{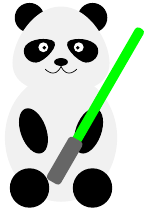}% 
\or
\includegraphics[height=2ex]{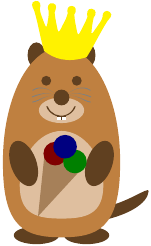}%
\or
\includegraphics[height=2ex]{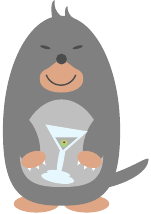}%
\or
\includegraphics[height=2ex]{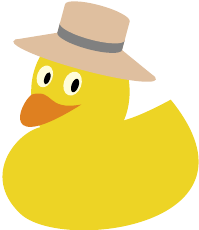}%
\or
\includegraphics[height=2ex]{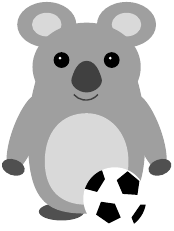}%
\or
\includegraphics[height=2ex]{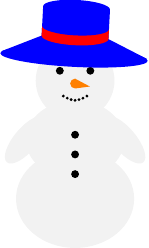}%
\or
\includegraphics[height=2ex]{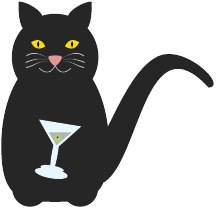}%
\or
\includegraphics[height=2ex]{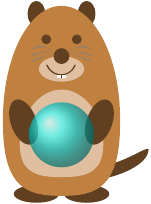}%
\or
\includegraphics[height=2ex]{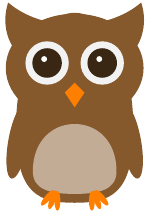}%
\or
\fi}
\makeatletter
\newcommand{\MyFootnote}[1][\@nil]{%
  \stepcounter{authorno}%
  \textsuperscript{\authorpic}%
  \def\tmp{#1}%
   \ifx\tmp\@nnil
    \else
       $^{,#1}$
    \fi}
\makeatother
\textbf{%
Yahya Almumin\rlap,\MyFootnote[a]%{yalmumin@uci.edu}
Mu--Chun Chen\rlap,\MyFootnote[a]%{muchunc@uci.edu}
Murong Cheng\rlap,\MyFootnote[b]%{murongc2@illinois.edu}
V\'ictor Knapp--P\'erez\rlap,\MyFootnote[a]%{vknapppe@uci.edu}
Yulun Li\rlap,\MyFootnote[c]%{yulunl@vt.edu}
Adreja Mondol\rlap,\MyFootnote[a]%{amondol@uci.edu}
Sa\'ul Ramos--S\'anchez\rlap,\MyFootnote[d]%{ramos@fisica.unam.mx}
Michael Ratz\MyFootnote[a]%{mratz@uci.edu} 
and 
Shreya Shukla\MyFootnote[a]%{sshukla4@uci.edu}
}
\setcounter{authorno}{0}
\begin{tikzpicture}[overlay,remember picture]
\path (current page.south) + (0,1) node[text width=1.1\linewidth,above]
{\MyFootnote{yalmumin@uci.edu},
\MyFootnote{muchunc@uci.edu},
\MyFootnote{murongc2@illinois.edu},
\MyFootnote{vknapppe@uci.edu},
\MyFootnote{yulunl@vt.edu},
\MyFootnote{amondol@uci.edu},
\MyFootnote{ramos@fisica.unam.mx},
\MyFootnote{mratz@uci.edu},
\MyFootnote{sshukla4@uci.edu}
};
\end{tikzpicture}
\\[8mm]
\textit{$^a$\small
~Department of Physics and Astronomy, University of California, Irvine, CA 92697-4575, USA
}
\\[5mm]
\textit{$^b$\small 
~Department of Physics, University of Illinois at Urbana-Champaign, Urbana, IL 61801, USA
}
\\[5mm]
\textit{$^c$\small 
~Department of Physics, Virginia Tech, Blacksburg, VA 24061, USA
}
\\[5mm]
\textit{$^d$\small ~Instituto de F\'isica, Universidad Nacional Aut\'onoma de M\'exico, POB 20-364, Cd.Mx. 01000, M\'exico}
\end{center}

\vspace*{1cm}

\begin{abstract}
The neutrino sector offers one of the most sensitive probes of new physics
beyond the \ac{SM}. The mechanism of neutrino mass generation is still unknown.
The observed suppression of neutrino masses hints at a large scale, conceivably
of the order of the scale of a \ac{GUT}, a unique feature of neutrinos that is
not shared by the charged fermions. The origin of neutrino masses and mixing is
part of the outstanding puzzle of fermion masses and mixings, which is not
explained in the \ac{SM}. Flavor model building for both quark and lepton
sectors is important in order to gain a better understanding of the origin of
the structure of mass hierarchy and flavor mixing, which constitute the dominant
fraction of the \ac{SM} parameters.

Recent activities in neutrino flavor model building based on non--Abelian
discrete flavor symmetries and modular flavor symmetries have been shown to be a
promising direction to explore. The emerging models provide a framework that has
a significantly reduced number of undetermined parameters in the flavor sector.
In addition, such framework affords a novel origin of \CP violation from group
theory, due to the intimate connection between physical \CP transformation and 
group theoretical properties of non--Abelian discrete groups.

Model building based on non--Abelian discrete flavor symmetries and their
modular variants enables the particle physics community to interpret the current
and anticipated upcoming data from neutrino experiments. Non--Abelian discrete
flavor symmetries and their modular variants can result from compactification of
a higher--dimensional theory. Pursuit of flavor model building based on such
frameworks thus also provides the connection to possible UV completions, in
particular to string theory. We emphasize  the importance of constructing models
in which the uncertainties of theoretical predictions are smaller than, or at
most compatible with, the error bars of measurements in neutrino experiments.
While there exist proof--of--principle versions of bottom--up models in which the
theoretical uncertainties are under control, it is remarkable that the key
ingredients of such constructions were discovered first in top--down model
building. We outline how a successful unification of bottom--up and top--down
ideas and techniques may guide us towards a new era of precision flavor model building
in which future experimental results can give us crucial insights in the UV
completion of the \ac{SM}.
\end{abstract}
\end{titlepage}
\renewcommand*{\thefootnote}{\arabic{footnote}}
\setcounter{footnote}{0}

% ======================================================================
\acresetall% 

\section{Introduction}

One of the most pressing questions in modern particle physics is what underlies
the \ac{SM}. While we continuously expand our understanding of how a consistent
theory of quantum gravity may look like, it is far less clear how the \ac{SM}
may fit into such a scheme. The lack of direct evidence of new physics at
current collider experiments appears to prevent us from inferring what a \ac{UV}
completion of the \ac{SM} may look like. On the other hand, the discovery of
neutrino oscillations has provided the very first compelling piece of
evidence for new physics beyond the \ac{SM}. While we have a rather good
comprehension of the structure of the \ac{SM} gauge sector, a fundamental
understanding of the structure of the flavor sector, which possesses the dominant
fraction of the \ac{SM} parameters, is still lacking. At present, the mechanism 
for neutrino mass generation is still unknown. Given that neutrinos are the only neutral 
fermions in the \ac{SM}, there exist many possible new physics scenarios that can 
yield masses for either Dirac or Majorana neutrinos. 
Given the expected wealth of experimental data obtained from current and future
neutrino experiments, it is imperative to construct robust flavor models, capable
of providing nontrivial testable predictions, in order to understand and interpret the data,
while at the same time allowing us to relate these predictions to properties of the physics that may
complete the \ac{SM} in the \ac{UV}.

Utilizing past and existing efforts to understand the origin of flavor, the purpose of this 
White Paper is to demonstrate that some of the most compelling bottom--up  
models have a clear connection to candidates for a consistent description of quantum 
gravity, as they possess common ingredients or employ similar approaches. These common 
features thus serve as examples of phenomenological applications of formal tools developed from  
top--down constructions aiming to \ac{UV} complete the \ac{SM}. On the other hand, efforts on 
bottom--up model building provide a way towards identifying those top--down constructions 
that are realized in nature.

More specifically, it has been known for more than 30 years that the
Yukawa couplings in certain types of string compactifications are modular forms
\cite{Ferrara:1989bc,Chun:1989se} (cf.\ the discussion around Equation~(19) of
\cite{Quevedo:1996sv}). However, only much more recently explicit neutrino mass
models utilizing modular forms have been put forward~\cite{Feruglio:2017spp}. In
the bottom--up approach, motivated by the observed large neutrino mixing, there 
have been many flavor models being proposed, utilizing non--Abelian discrete
groups~\cite{Kaplan:1993ej}, which we
review in \Cref{sec:TraditionalFlavorSymmetries}. Despite major efforts over
many years, it is probably fair to say that this approach has not yet provided
us with a complete and compelling picture. A big obstacle in making this scheme
fully successful is the fact that these symmetries need to be broken, and
breaking the symmetries typically lead to ad hoc choices and additional
parameters which limit the predictive power of the scheme. As we will discuss in
more detail in \Cref{sec:ModularFlavorSymmetries}, Feruglio's
approach~\cite{Feruglio:2017spp} avoids these complications largely. 

This paper is organized as follows. In \Cref{sec:LeptonData} we briefly review
the current status of our understanding of the lepton sector, and the expected
outcomes of current and future neutrino experiments. In \Cref{sec:MassGen} we
discuss various mechanisms that have been proposed to explain neutrino mass
generation. Some aspects of discrete symmetries in flavor physics are reviewed
in \Cref{sec:TraditionalFlavorSymmetries}. Their modular cousins are discussed in
\Cref{sec:ModularFlavorSymmetries}. \Cref{sec:Summary} contains a discussion and
outlook.

\section{What do we know about the lepton sector?}
\label{sec:LeptonData}

\subsection{What do we currently know?}

About half a century ago, the possibility of massive neutrinos was theoretically
introduced, leading to the notion of neutrino oscillations, described  by the
Pontecorvo--Maki--Nakagawa--Sakata (PMNS) mixing 
matrix~\cite{Pontecorvo:1957qd,Gribov:1968kq,Maki:1962mu}. Super--Kamiokande and
the Sudbury Neutrino Observatory gave a direct evidence of atmospheric neutrino
oscillations~\cite{Super-Kamiokande:1998kpq} and solar neutrino
oscillations~\cite{SNO:2001kpb}, respectively, providing the very first 
evidence of physics beyond the \ac{SM}. In a little more than two decades, our
community  went from seeing the first evidence of nonvanishing neutrino mass to
measuring the three  mixing angles and two squared mass differences with good
precision. In addition, with the  wealth of experimental data available, we have
now some hint for a nonvanishing Dirac  \CP phase $\delta_{\CP}$. The results
from the global fit of the mixing angles and mass splittings are summarized in
\Cref{tab:CurrentData}. 

While the measurements of several neutrino oscillation parameters have entered
a precision era, there are still many outstanding puzzles in regards to
neutrino properties. First of all, oscillation experiments can only inform us
of the squared mass differences instead of the absolute masses. Current
experimental data is still consistent with the neutrino mass spectrum either
with the two lightest neutrinos having a smaller  mass difference, which is
defined as the \ac{NO}, or with the two lightest neutrinos  having a larger mass
difference, which is defined as the \ac{IO}. For both \ac{NO} and \ac{IO}, 
$\Delta m_{21}^2=(7.42^{+0.21}_{-0.20})\times 10^{-5}\,\text{eV}^2$,  but the
second squared mass difference depends on the ordering. For \ac{NO}, $\Delta
m_{31}^2=(2.51\pm0.027)\times 10^{-3}\,\text{eV}^2$, and for \ac{IO}, $\Delta
m_{32}^2=(-2.49^{+0.026}_{-0.028})\times10^{-3}\,\text{eV}^2$~\cite{Esteban:2020cvm,NuFit:2021}.
On the other hand, there still exists a tension between the best fit values of
the solar parameters $\theta_{12}$ and $\Delta m_{21}^{2}$ in KamLAND and the
solar neutrino analysis, even though it has been reduced from $2.2\sigma$ to
$1.1\sigma$~\cite{Esteban:2020cvm}. For the atmospheric parameters,
$\theta_{23}$ and $\Delta m_{31}^{2}$, the \ac{IO} fit is more consistent
between T2K and NOvA than the \ac{NO} fit~\cite{Esteban:2020cvm}. In addition,
for the Dirac \CP phase $\delta_{\CP}$, T2K gives the current best fit value, 
$195^{\circ}~$\cite{T2K:2019bcf}, which is closer to the \CP conserving value,
$180^{\circ}$ ($0.6\sigma$), than it was in the previous results,
$215^{\circ}$~\cite{Esteban:2018azc}.

\begin{table}[t!]
 \centering
 \includegraphics[width=0.85\textwidth]{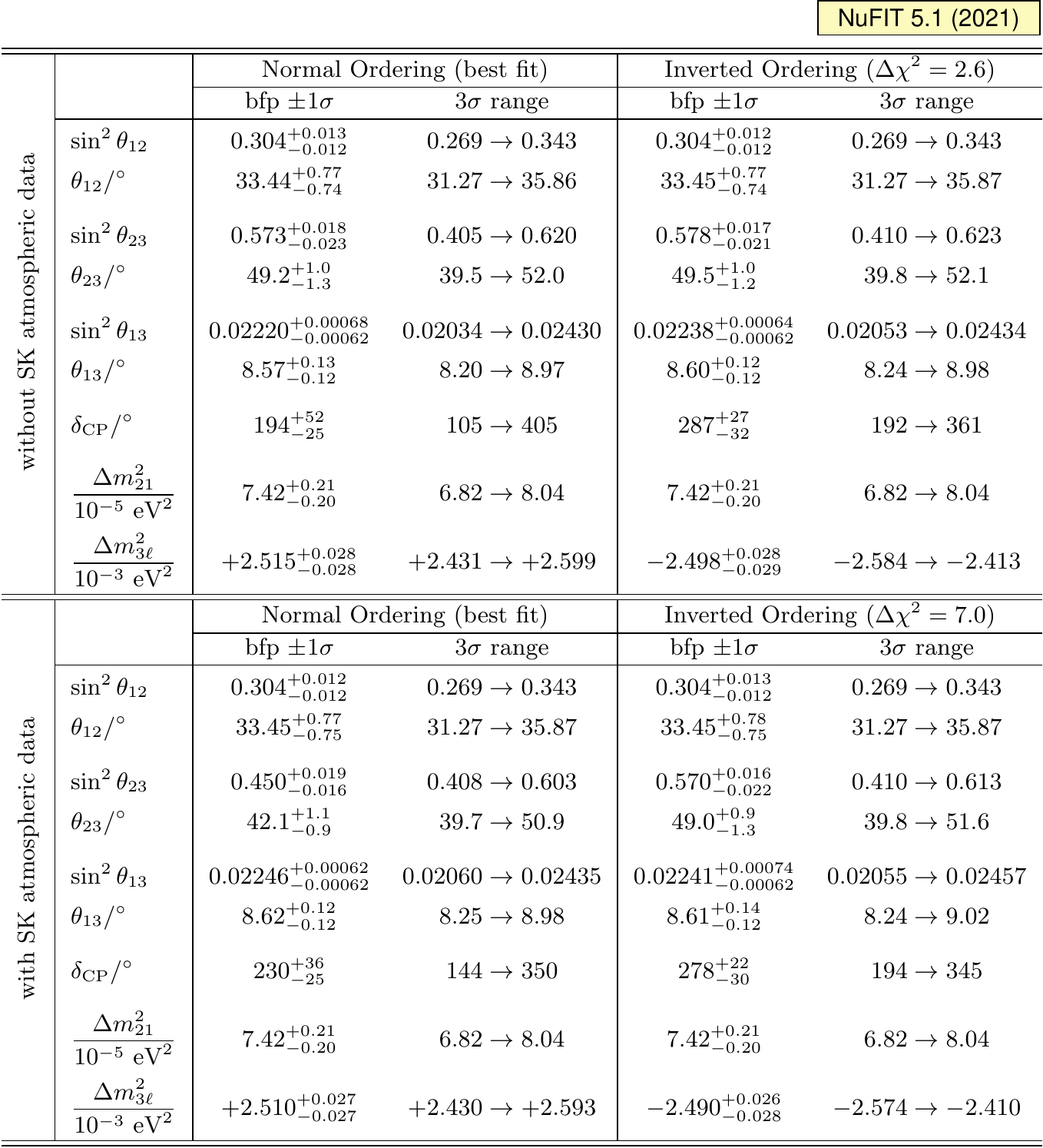}
 \caption{Current best--fit values of the leptonic mixing parameters. 
 	Taken from~\cite{Esteban:2020cvm,NuFit:2021}.}
 \label{tab:CurrentData}
\end{table}

Different types of experiments have been built to further reduce the
uncertainties in the measurements. It is conceivable that there might be
tensions among  different experiments. These potential tensions thus could be a
pathway to unravel some new physics~\cite{Chatterjee:2022nia}. Interesting
neutrino experiments that might unveil new directions in physics include the
scattering of  high energy neutrinos against different targets. For example,
$\nu_e-e$ scattering or coherent elastic $\nu-$nucleus scattering (CE$\nu$NS)
have successfully refined the limits of current measurements and  tested some
proposed new physics scenarios~\cite{NOvA:2021nfi}.

\subsection{What do we expect to know?}

\begin{figure}[t!]
 \centering
 \includegraphics[width=0.6\textwidth]{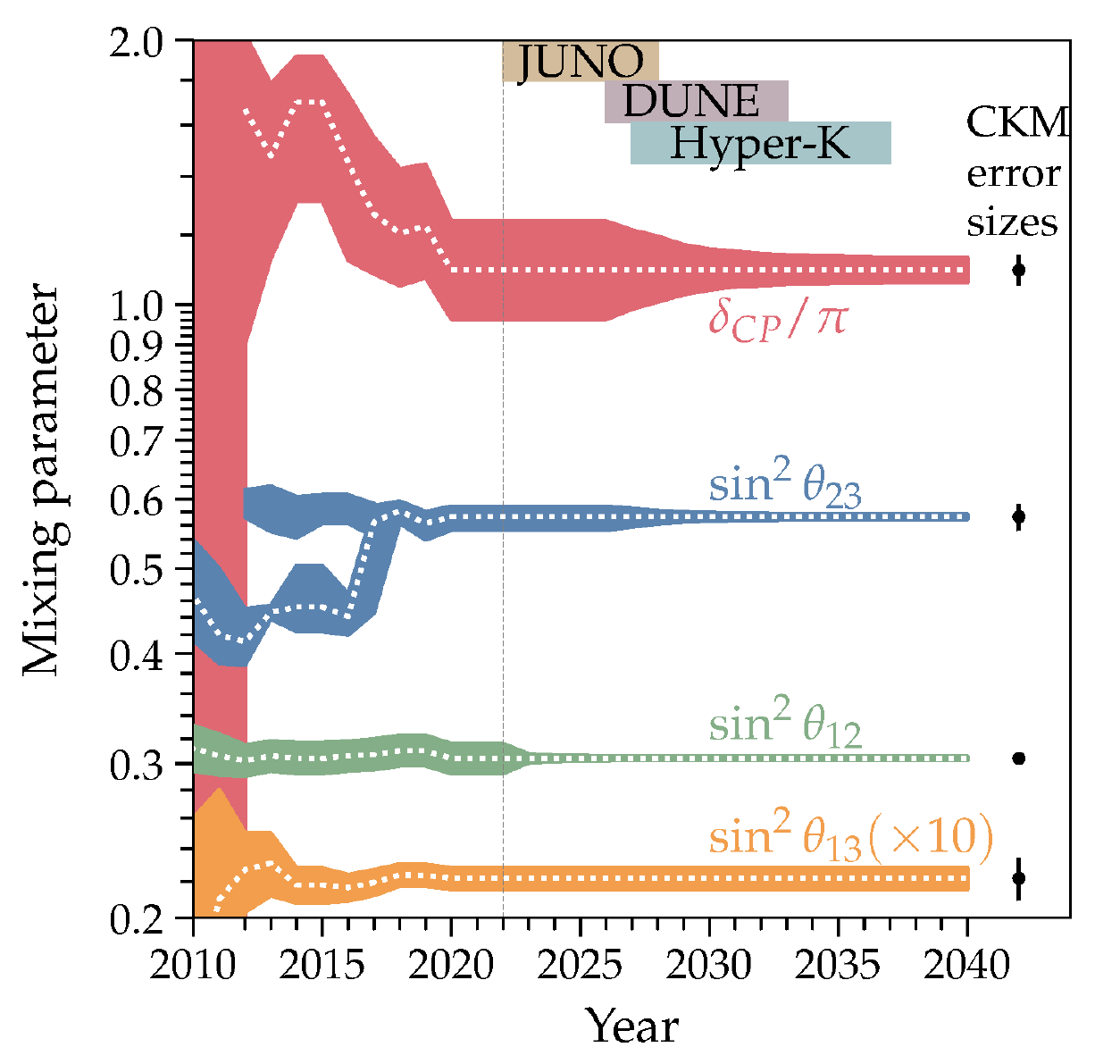}
 \caption{Current and expected  precision of measurements for leptonic 
  parameters in three--neutrino flavor framework \cite{Song:2020nfh}. Courtesy
  of Shirley Li.}
 \label{fig:ExpectedPrecision}
\end{figure}

Current and future experiments will allow us to pin down the leptonic mixing
parameters with a precision that is comparable to, or even better than, the one
in the quark sector (see \Cref{fig:ExpectedPrecision}). In the near future, the
currently running experiments T2K and NOvA will not significantly further 
improve the precision on $\delta_{\CP}$. However, there are long--baseline
neutrino experiments under construction designed to investigate the leptonic \CP
violation. We are expecting to see \CP violation at $3\sigma$ significance for
75\% of the $\delta_{\CP}$ allowed range in Hyper--Kamiokande's 10 years of
operation~\cite{Hyper-Kamiokande:2018ofw}. Also, DUNE is expected to get
sensitivity at $3\sigma$ significance for more than 75\% in 14 years of
operation~\cite{DUNE:2016hlj}. Besides the measurements in beam experiments,
$\delta_{\CP}$ is also expected to be measured by JUNO, which is also under
construction. They plan to use the combination of JUNO detector and a
superconductive cyclotron to get $3\sigma$ significance for 22\% of the
$\delta_{\CP}$ allowed range~\cite{Smirnov:2018ywm}.

Depending on which mass ordering Nature chooses, the physics predicted from them
could be different. For instance, the neutrinoless double beta decay depends on
a mass term $\langle m_{\beta\beta} \rangle$ which is related to the lightest
mass eigenstate. If Nature follows \ac{NO}, then the lightest mass will be
$m_1$, which will urge the experimental resolution to be better than
$0.004\,\text{eV}$ in order to measure the neutrinoless double beta decay.
However, our current experiments cannot achieve this precision
yet~\cite{Denton:2019ovn}.

On the other hand,  present and future experiments, including Super--Kamiokande,
NOvA, DUNE, JUNO, etc., are working on determining the mass ordering. For
instance, Super--Kamiokande is using atmospheric neutrinos by comparing $\Delta
m_{32}^2$ electron and muon neutrino disappearance channels. If \ac{NO} is
favored, then the squared mass difference in electron neutrino disappearance
will be larger than that in muon neutrino disappearance, and vice versa for
\ac{IO}~\cite{Qian:2015waa}. 

There is also physics beyond the \ac{SM} that we are expecting to learn soon.
This includes higher--order interference~\cite{Lee_2016}, and neutrino
decoherence~\cite{Xu:2020pzr}. There is an advantage of probing the Sorkin's 
triple path interference with neutrino oscillation in JUNO. The accuracy of 
probing for triple--path interference in JUNO with neutrinos is comparable
to that of the electromagnetic probes.
%\todo{\MR{I removed two sentences here.}} 
% In addition, closing and opening the
% slits  on the electromagnetic probes set a limit on improving the accuracy of 
% electromagnetic probes. On the contrary, neutrino oscillation is independent  of
% this condition~\cite{Huber:2021xpx}.\todo{\SRS{Which condition? Sorry, I don't
% understand the last sentence.}}

\subsection{What do we want to know?}

At the same time, we currently do not even know which operators should be
included in the Lagrange density of the \ac{SM} in order to provide the correct
description of the mechanism that generates neutrino masses. That is, we do not
know whether neutrinos are Dirac or Majorana particles, nor do we know what the
scale of neutrino mass generation is. In the case of Majorana neutrinos, there
are two more parameters, the so--called Majorana phases, about which we do not
have any experimental information. 

According to the review paper~\cite{Agostini:2022zub}, 0$\nu\beta\beta$ decay
experiments done by KamLAND--Zen and cosmological observations done by
Planck space observatory set a limit to the absolute neutrino mass. In the next
generation experiments, the lower scale 0$\nu\beta\beta$ experiments, the ECHo
experiments, tritium $\beta$ decay experiments, and the \ac{CnB} experiments will 
be likely to determine the absolute neutrino mass
scale~\cite{Cirigliano:2022oqy,Gastaldo:2017edk,PTOLEMY:2019hkd,Project8:2017nal}. 
They can reach the sensitivity in the sub--eV region, and in particular,
the tritium $\beta$ decay can reach $\sim 0.04\,\text{eV}$. The PTOLEMY project 
is proposed to develop a \ac{CnB} detector to search for cosmic neutrinos with a 
sensitivity dependent on the absolute neutrino mass scale~\cite{PTOLEMY:2019hkd}.

The anomalies arising in various experiments, e.g., LSND, T2K and NOvA,  also
hint at the possible existence of new physics. There are several popular
phenomenological frameworks that go beyond the standard  framework with three
neutrino flavors, including scenarios with \ac{NSI}, with dark or  sterile
neutrinos, and with additional light scalars or light vectors (such as light
$Z'$ or dark photons). \ac{NSI} assume that neutrinos can be coupled to the
charged leptons of  different flavors. The sterile neutrino framework, on the
other hand, introduce new neutrino species that are not coupled through the weak
interactions. Both the \ac{NSI} and sterile neutrinos scenarios are often
utilized  as a solution to explain  the experimental
anomalies~\cite{Danilov:2022str}. Neutrino scattering experiments such as the
Dresden--II reactor and the COHERENT experiments can probe new physics
scenarios~\cite{Coloma:2022avw}, such as the dark photon, light scalar, or light
vector framework.

Having reviewed the current experimental status and the expected sensitivity
of future experiments, we will next turn the theoretical description of neutrino
mass terms in \Cref{sec:MassGen}.

\section{Neutrino mass generation}
\label{sec:MassGen}

As we have seen in \Cref{sec:LeptonData}, neutrino masses are quite different
from the masses of the other fermions in the \ac{SM}. In particular,
\begin{enumerate}
 \item neutrinos are substantially lighter than even the lightest charged
  fermions, and
 \item leptonic mixing angles are generally much larger than their counterparts
  in the quark sector.
\end{enumerate}
As mentioned previously, we do not know at present what operators in the 
Lagrange density are responsible for neutrino mass generation. Given that neutrinos 
are the only neutral fermions in the \ac{SM}, there are more ways for 
neutrinos to acquire masses than for charged leptons.

In the following we will briefly summarize some aspects of various mechanisms
that have been proposed for neutrino mass generation. Depending on whether 
neutrinos are Majorana or Dirac fermions, these mass generation mechanisms based
on  a variety of new physics frameworks differ in the new particles and
symmetries introduced,  the scales at which the mechanisms take place, and thus
the ways neutrino masses are suppressed. In this Section we focus on the
question why neutrino masses are suppressed, leaving the flavor aspects mainly
to \Cref{sec:TraditionalFlavorSymmetries,sec:ModularFlavorSymmetries}.

\subsection{Mass generation for neutrinos as Majorana fermions}
\label{sec:Maj}

\begin{figure}[h!]
   \centering
      \begin{subfigure}[b]{.3\linewidth}
        \centering\includegraphics{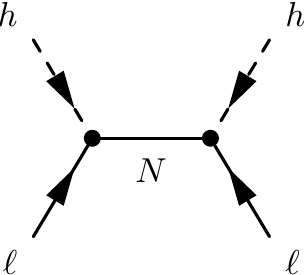}
        \caption{Type I.}\label{fig:Seesaw-a}
      \end{subfigure}%
      \begin{subfigure}[b]{.3\linewidth}
        \centering\includegraphics{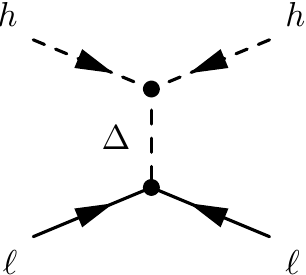}
        \caption{Type II.}\label{fig:Seesaw-b}
      \end{subfigure}%
      \begin{subfigure}[b]{.3\linewidth}
        \centering\includegraphics{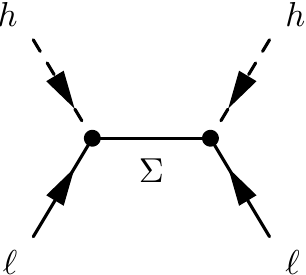}
        \caption{Type III.}\label{fig:Seesaw-b}
      \end{subfigure}%
	  \caption{Seesaw diagrams. The canonical seesaw mechanism is also referred
	   to as type I. Here $N$ is a right--handed neutrino whereas $\Delta$ and
	   $\Sigma$ are scalar and fermionic $\text{SU}(2)_\text{L}$ triplets,
	   respectively.}
	  \label{fig:Seesaw}
\end{figure}

A particularly compelling scheme to address the smallness of neutrino masses is
the seesaw mechanism
\cite{Minkowski:1977sc,Yanagida:1979as,Glashow:1979nm,Gell-Mann:1979vob,Magg:1980ut,Lazarides:1980rn,Mohapatra:1979ia,Mohapatra:1980yp,Foot:1988aq}. 
In its simplest form
\cite{Minkowski:1977sc,Yanagida:1979as,Glashow:1979nm,Gell-Mann:1979vob,Lazarides:1980rn,Mohapatra:1979ia},
right--handed neutrinos are introduced. The existence of such right--handed
neutrinos is motivated by the scheme of \acp{GUT}, in particular the
\SO{10} model~\cite{Fritzsch:1974nn}, where right--handed neutrinos are unavoidably 
predicted. Integrating them out leads to the Weinberg operator 
\begin{equation}\label{eq:Lkappa}
 \mathscr{L}_\kappa=\frac{1}{4}\kappa_{gf}
 \overline{\ell^{\ChargeC,g}}h\,\ell^f h
 +\text{h.c.}\;,
\end{equation} where $\ell^f$ denote the left--handed lepton doublets of the
\ac{SM},  $h$ the Higgs doublet, and $g$ and $f$ are the family indices. 
\eqref{eq:Lkappa} is the unique dimension--five operator consistent with the
symmetries of the \ac{SM}. In the canonical seesaw mechanism, $\kappa_{gf}$
scales like the inverse of the masses of the right--handed neutrinos $M$, and
the smallness of the active neutrino mass eigenvalues $m_\nu$ is explained via
the famous seesaw formula
\begin{equation}\label{eq:SeesawFormula}
 m_\nu\simeq\frac{(m_\nu^\text{Dirac})^2}{M}\;.
\end{equation}
Here, the Dirac neutrino mass, $m_\nu^\text{Dirac}$, is given by the product  of
the Dirac neutrino Yukawa matrix $Y_\nu$ and the Higgs \ac{VEV}. Of course,
\eqref{eq:SeesawFormula} is to be understood as a shorthand for a matrix
equation. There are variations of the canonical seesaw, known as type II
\cite{Magg:1980ut,Lazarides:1980rn,Mohapatra:1979ia} and type III
\cite{Mohapatra:1980yp,Foot:1988aq}, in which heavy $\SU2_\mathrm{L}$ triplets
get exchanged. The different variants are depicted in \Cref{fig:Seesaw}. 

\subsection{Radiative neutrino masses}

The underlying idea of the so--called radiative neutrino mass models 
(see e.g.\ \cite{Cai:2017jrq} for a review) is that
neutrino masses, which are absent at the tree level, may get induced via loops,
thus addressing the smallness of neutrino masses. This option was pioneered by
Zee~\cite{Zee:1980ai}.

\begin{wrapfigure}[12]{r}[10pt]{6.5cm}
 \centering
 \includegraphics{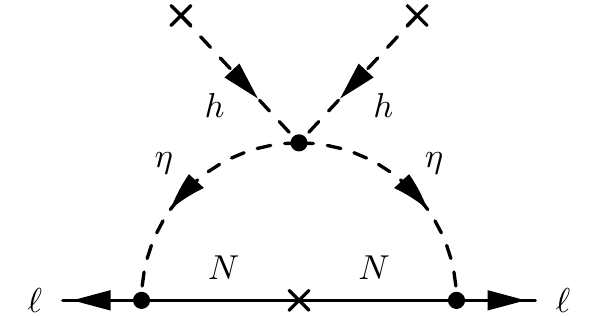}
 \caption{One--loop diagram inducing neutrino masses. Recreated from 
 	\cite[Section~5.3]{Cai:2017jrq}.}
 \label{fig:ScotogenicLoop}
\end{wrapfigure}
Here we will use the so--called scotogenic model~\cite{Ma:2006km} to review 
some of the relevant facts. A nice overview for this model can be found 
in~\cite[Section~5.3]{Cai:2017jrq}. This model contains the \ac{SM} fermions, 
three singlet fermions $N^f$ as well as an extra doublet $\eta$ which
has the same \ac{SM} quantum numbers as the \ac{SM} Higgs doublet $h$ but is
distinguished by an extra $\Z2$ symmetry. In more detail, both the $N^f$ as well
as $\eta$ are odd under this $\Z2$ whereas the usual \ac{SM} fields are even.
This $\Z2$ forbids Yukawa couplings between $\ell$, $h$ and $N$ but allows for
couplings between $\ell$, $\eta$ and $N$. However, there are interactions
between $h$ and $\eta$ via a scalar potential, and these interactions give rise
to a loop diagram (cf.\ \Cref{fig:ScotogenicLoop}) which induces an effective
Weinberg operator, and thus neutrino masses. 

Note that this diagram requires a coupling $\re(h^\dagger\eta)^2$. As explained
in \cite[Section~5.3]{Cai:2017jrq}, this term breaks a ``generalized lepton
number'' symmetry that the Lagrange density would otherwise have, which forbids
(Majorana) neutrino masses. A point that will be relevant for our later
discussion in \Cref{sec:TraditionalFlavorSymmetries,sec:ModularFlavorSymmetries}
is that this model does, in the presence of the above--mentioned quartic
coupling, not have a symmetry that forbids the Weinberg operator. So one could a
priori just add it to the model. However, all the terms required for the model
to work are renormalizable. Therefore, even if a tree--level Weinberg operator
exists, for a large enough cut--off scale $\Lambda$ its contribution to the
light neutrino masses can be suppressed against the loop term. In this sense,
the neutrino masses are generated by loops even though other contributions
are not completely  forbidden. 

Thus, this \emph{renormalizable} model can provide us with robust relations
between the model parameters and neutrino data. Later, in
\Cref{sec:TraditionalFlavorSymmetries,sec:ModularFlavorSymmetries} we will see 
that extra contributions cannot always be sufficiently suppressed in
constructions which rely on higher--dimensional operators without explicit
\ac{UV} completion.

\subsection{Dirac neutrino masses}

The charged \ac{SM} fermions get their masses from Dirac mass terms, which
combine two different Weyl spinors. This option is also valid for neutrinos,
for which these mass terms then originate from the Yukawa terms
\begin{equation} \label{eq:nu_D}
 \mathscr{L}_\nu^\text{Dirac}= Y_\nu^{gf} \overline{\ell^f}h N^g  + \text{h.c.}\;,
\end{equation}
where $N^g$ denotes the right--handed neutrinos.  While small fermion masses are
said to be technically natural, the required Yukawa couplings
$|Y_\nu^{gf}|\lesssim10^{-12}$ beg for an explanation. Rather compelling
mechanisms have been put forward in the past. Many of these mechanisms rely on
ingredients in new physics scenarios that aim at solving the gauge hierarchy
problem. These include supersymmetry~\cite{Arkani-Hamed:2000oup}, radiative
mechanisms~\cite{Babu:1988yq,Farzan:2012ev}, warped extra 
dimensions~\cite{Grossman:1999ra,Huber:2000ie}, and more recently the 
clockwork mechanism~\cite{Park:2017yrn,Hong:2019bki,Babu:2020tnf}. 
Dirac neutrino masses may also just arise at higher orders~\cite{Babu:2001ex}. 
While these possibilities can be argued to deserve more attention, in the following, 
we restrict our focus on seesaw scenarios for Majorana neutrinos.

\subsection{Neutrino masses in explicit string models}

A long--standing question is what string theory says about neutrino masses 
\cite{Giedt:2005vx}. The answer is model--dependent. It has been proposed that
the smallness of neutrino masses may be due to their origin from instantons
\cite{Blumenhagen:2006xt}. However, an explicit  \ac{SM}--like model featuring
this scenario has yet to be found. On the other hand, the heterotic string
provides us with an abundance of explicit \ac{SM}--like models with
seesaw--suppressed neutrino masses~\cite{Buchmuller:2007zd}. Importantly, in
string models the spectrum is fixed, and one can simply count the number of
right--handed neutrinos. It turns out that, unlike in bottom--up models, their
number is of the order $N_\nu=\mathcal{O}(10\dots100)$. This means that the
Weinberg operator~\eqref{eq:Lkappa} receives contributions from many neutrinos,
\begin{equation}
 \vcenter{\hbox{\includegraphics{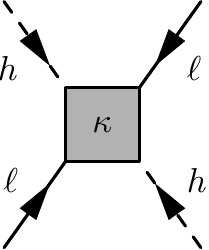}}}
 =
 \sum_{f=1}^{N_\nu}
 \vcenter{\hbox{\includegraphics{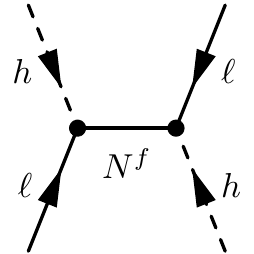}}}
 +
 \vcenter{\hbox{\includegraphics{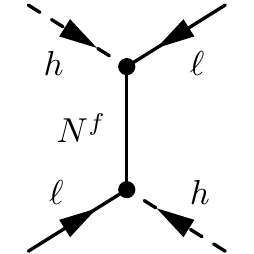}}}\;.
\end{equation}
This effectively lowers the seesaw scale, i.e.\ even if the individual
right--handed neutrino mass eigenvalues are of the order of the \ac{GUT} 
scale $\mathcal{O}(10^{16})\, \text{GeV}$, 
realistic active neutrino
masses can emerge. It has further been found in~\cite{Feldstein:2011ck} that the
contribution of many neutrinos can mimic the  anarchy scenario
\cite{Hall:1999sn,deGouvea:2003xe,deGouvea:2012ac}, in which large mixing angles
are statistically favored. 

On the other hand, in string models one cannot only count the right--handed
neutrinos, one can, at least in principle, compute the couplings. These
couplings are often constrained by ``traditional'' flavor symmetries, which we
will study in \Cref{sec:TraditionalFlavorSymmetries}. Moreover, additional
constraints arise from modular symmetries, which are  intrinsic to string
theory; we will discuss these symmetries, their origin and their role in
phenomenology in \Cref{sec:ModularFlavorSymmetries}. That is, in what follows,
we will  concentrate on an alternate, i.e.\ non--anarchic, approach to
understand the origin of flavor. Specifically, we will discuss how the observed
large mixing angles may arise from the dynamics of certain underlying
fundamental symmetries.

\section{Traditional flavor symmetries}
\label{sec:TraditionalFlavorSymmetries}

A curious feature of the \ac{SM} is the repetition of families, i.e.\ the fact
that \ac{SM} matter appears in three copies of particles transforming in
identical representations under the \ac{SM} gauge group,
\begin{equation}
 G_\text{SM}=\SU3_C\times\SU2_\text{L}\times\U1_Y\;. 
\end{equation}
This repetition may be the consequence of a so--called horizontal or flavor
symmetry.
Continuous flavor symmetries often suffer from anomalies and/or unrealistic
Goldstone modes, so a lot of attention has been given to non--Abelian finite
groups~\cite{Kaplan:1993ej} (see e.g.~\cite{Ishimori:2010au} for a review). These symmetries are less 
challenged by Goldstone modes.\footnote{The anomalies of finite groups can readily be
determined~\cite{Araki:2006sqx,Araki:2008ek,Chen:2015aba,Talbert:2018nkq,Kobayashi:2021xfs,Gripaios:2022vvc}, 
yet their implications have not been worked in great detail so far in the context of
(bottom--up) model building. Discrete matching~\cite{Csaki:1997aw} of these anomalies 
as well as outer automorphism anomalies~\cite{Henning:2021ctv} may provide us with 
crucial insights on how bottom--up and top--down models are related.}  
Imposing such flavor symmetries can help to reduce number of free
parameters, and lead to nontrivial predictions. It is not the purpose of this
paper to survey all flavor models (for reviews and references see e.g.\
\cite{Ishimori:2010au,Feruglio:2019ybq}). Rather, in what follows we will illustrate 
some of the main prospects and challenges using a concrete example.

\subsection{Example: $A_4$}
\label{sec:ExampleA4}

\subsubsection{Explicit model}
\label{sec:ExplicitA4Model}

One particularly popular example of a finite flavor symmetry for the lepton
sector of the \ac{SM} is the alternating group of order 4, $A_4$
\cite{Ma:2001dn,Babu:2002dz,Hirsch:2003dr,Altarelli:2005yp}. An explicit
$A_4$--based example assumes low--energy \ac{SUSY}, and the relevant
superpotential terms are given by~\cite{Altarelli:2005yp}
\begin{subequations}\label{eq:A4_model_superpotential}
\begin{align}
\mathscr{W}_\nu
 & =
 \frac{\lambda_1}{\Lambda\,\Lambda_\nu}
 \left\{\left[\left(L\Hu\right) 
 	\otimes
	\left(L\Hu\right)
	\right]_{\rep{3}} 
 \otimes \AfourFlavonA\right\}_{\rep{1}}
 +
 \frac{\lambda_2}{\Lambda\,\Lambda_\nu}
 \left[
 	\left(L\Hu\right)
	\otimes
	\left(L\Hu\right)
 \right]_{\rep{1}}\,
 \xi\;,\\
 \mathscr{W}_e
 & = 
 \frac{h_e}{\Lambda}
 \left(\AfourFlavonB\otimes L\right)_{\rep{1}}
 \,\Hd\,e_\mathrm{R}
 +
 \frac{h_\mu}{\Lambda}\left(\AfourFlavonB\otimes
 L\right)_{\rep{1'}}\,\Hd\,
 \mu_\mathrm{R}
 +
 \frac{h_\tau}{\Lambda}\left(\AfourFlavonB\otimes
 L\right)_{\rep{1''}}
 \,\Hd\,\tau_\mathrm{R}\;.
\end{align}
\end{subequations}
Here, $L$, $e_\mathrm{R}$, $\mu_\mathrm{R}$, $\tau_\mathrm{R}$, $\Hu$, $\Hd$, 
respectively, denote the lepton doublets, charged leptons, $u$--type Higgs,
$d$--type Higgs   of the (minimal) supersymmetric \ac{SM}. The lepton doublets,
$L=(L_e,L_\mu,L_\tau)^T$, are assumed to transform as an $A_4$ \rep{3}--plet.
The right--handed charged lepton fields, $\mu_\mathrm{R}$ and $\tau_\mathrm{R}$
transform in the one--dimensional representations $\rep{1''}$ and $\rep{1'}$ of
$A_4$, respectively. The Higgs doublets and $e_\mathrm{R}$ are trivial $A_4$ 
singlets. The fields $\AfourFlavonA$, $\AfourFlavonB$ and $\xi$ are
so--called flavons, and assumed to transform as \rep{3}, \rep{3} and \rep{1},
respectively. $\Lambda$ denotes the cut--off and $\Lambda_\nu$ the seesaw scale.
The subscripts \rep{1}, \rep{3} and so on denote contractions to a
\rep{1}--plet, \rep{3}--plet, etc. 

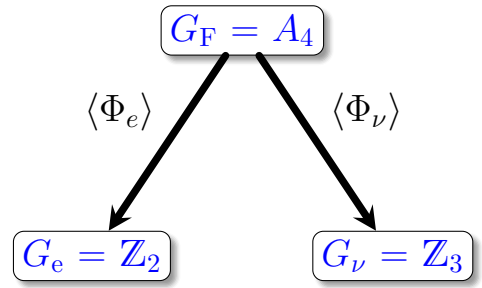
\begin{wrapfigure}[10]{r}[10pt]{7cm}
\centering
\begin{tikzpicture}[>=stealth,line cap=round]
\path[nodes={blur shadow,draw,rounded corners,text=blue,font=\Large,fill=white}] 
	(3,3.5) node(GF){$G_\mathrm{F}=A_4$}
	(1,0.5)	node(GL){$G_\mathrm{e}=\Z2$}
	(5,0.5) node(Gnu) {$G_\nu=\Z3$};
\draw[line width=1mm,->](GF)--(GL)
	node[midway,above left,font=\Large]{$\langle\AfourFlavonB\rangle$};
\draw[line width=1mm,->](GF)--(Gnu)
	node[midway,above right,font=\Large]{$\langle\AfourFlavonA\rangle$};	
\end{tikzpicture}
\caption{Approximate partial breaking of $A_4$.}
\label{fig:ApproximatePartialBreakingOfA4}
\end{wrapfigure}
The flavons are assumed to attain certain \acp{VEV},
\begin{subequations}\label{eq:VEValignmentA4}
\begin{align}
\langle\AfourFlavonA\rangle & = \left(v,v,v\right) \;,\\
\langle\AfourFlavonB\rangle & = \left(v',0,0\right) \;,\\
\langle\xi\rangle & =  w\;,
\end{align}
\end{subequations}
where $v$, $v'$ and $w$ are some dimensionful parameters that are to be
explained through a suitable mechanism that aligns the \acp{VEV}, a point that
we will expand on below in \Cref{sec:VEV_alignment}.
To first approximations, these \acp{VEV} break $A_4$ down to $G_e=\Z2$ in the
charged lepton sector and to $G_\nu=\Z3$ in the neutrino sector, cf.\
\Cref{fig:ApproximatePartialBreakingOfA4}. These approximate symmetries fix the
neutrino superpotential to be given by \cite{Altarelli:2005yp}
\begin{align}
 \mathscr{W}_\nu
 & = 
 \left(L_e\,\Hu,L_\mu\,\Hu,L_\tau\,\Hu\right)\,
 \left(\begin{array}{ccc}
                a + 2d & -d & -d \\
		-d & 2d & a -d \\
                -d & a - d
				& 2d
 \end{array}\right) \,
 \begin{pmatrix}
	L_e\,\Hu\\ L_\mu\,\Hu\\ L_\tau\,\Hu
 \end{pmatrix}	
 \label{eq:Wnu_A4}
\end{align} 
up to corrections that we will discuss below in
\Cref{sec:TraditionalCorrections}. The entries of the mass matrix in
\Cref{eq:Wnu_A4} depend, at leading order, only on the parameters
$a=2 \lambda_{1}\, \lambda_{2}\, \frac{w}{\Lambda}\,\frac{1}{\Lambda_\nu}$
and $d=\frac{\lambda_{1}}{3}\, \frac{v}{\Lambda}\,\frac{1}{\Lambda_\nu}$.
At the same time, the charged lepton superpotential leads to diagonal Yukawa
couplings,
\begin{align}
 \mathscr{W}_e
 & = 
 \left(L_e,L_\mu,L_\tau\right)\,
 \begin{pmatrix}
 y_e & 0 & 0\\ 0  & y_\mu & 0\\  0 & 0 & 
 y_\tau
 \end{pmatrix}\,
 \begin{pmatrix}e_\mathrm{R}\\ \mu_\mathrm{R}\\ \tau_\mathrm{R}\end{pmatrix}\,\Hd
 \;,
\end{align}
where $y_{e,\,\mu,\,\tau}=h_{e,\,\mu,\,\tau}\,\frac{v'}{\Lambda}$. As a 
result, the mixing is of the so--called tribimaximal form~\cite{Harrison:2002er}, 
i.e.\ the PMNS~\cite{Maki:1962mu} matrix is given by
\begin{equation}
 U_\mathrm{PMNS}^\mathrm{TBM}
 =
 \begin{pmatrix}
                \sqrt{\frac{2}{3}} & \frac{1}{\sqrt{3}} & 0 \\
		- \frac{1}{\sqrt{6}} & \frac{1}{\sqrt{3}} & - \frac{1}{\sqrt{2}} \\
	      - \frac{1}{\sqrt{6}} & \frac{1}{\sqrt{3}} &  \frac{1}{\sqrt{2}}
 \end{pmatrix}\;.
\end{equation}
This corresponds to the leptonic mixing angles
\begin{equation}
\theta_{12}\simeq35^\circ\;,\quad
\theta_{13}=0\quad\text{and}\quad
\theta_{23}=45^\circ\;.
\end{equation}
Even though these very angles are no longer consistent with data, they could be
regarded as a step towards a realistic mixing model.

The main point to make in the context of this model is that non--Abelian flavor
symmetries may provide us with predictions of the mixing angles that are, to
some approximation, independent of the continuous parameters of the model.

However, given the shear number of models which appear to describe observation
one may wonder how robust the predictions really are. In other words, if many
different symmetries are consistent with the data we have, to which extent do
these symmetries really predict the observed pattern of masses and mixings? In
what follows we shall review some of the limitations shared by many models based
on finite groups.

\subsubsection{Corrections and limitations}
\label{sec:TraditionalCorrections}

As we have seen in \Cref{sec:LeptonData}, specifically
\Cref{tab:CurrentData,fig:ExpectedPrecision}, our experimental community has
determined the neutrino parameters with an impressive accuracy, which will even
be dramatically improved further in the near future. It is, therefore, natural
to ask what the error bars in the predictions from the models are. It turns out
that often the error bars do not get specified. However, in the (bottom--up)
models there are often substantial uncertainties. It is rather easy to see why
this is. These models are usually \acp{EFT}, i.e.\ defined by the symmetries and
particle content, and endowed with a cut--off, $\Lambda$, as is the case in our
example in \eqref{eq:A4_model_superpotential}. The symmetries get spontaneously
broken by some flavon \acp{VEV}, and one finds rather modest hierarchies in
explicit flavor models, often the ratio of a typical flavon \ac{VEV} over the
cut--off scale is of the order of the Cabbibo angle,
$\varepsilon:=\Braket{\Phi}/\Lambda\sim0.2$. The best one can do in an \ac{EFT}
framework is to perform an expansion in $\varepsilon$, and these qualitative
arguments suggest that there are order $20\%$ corrections to predictions.

However, qualitative arguments do not always lead to the correct conclusions, so
one may wonder if these corrections arise, and if so, how. It turns out that in
supersymmetric theories higher order terms in the superpotential can often be
forbidden by some carefully crafted $R$ and non--$R$ symmetries. Likewise,
quantum corrections from the \acp{RGE} have been worked out analytically (cf.\
e.g.\ \cite{Antusch:2005gp}) and, while there is still room for relevant
corrections in the \ac{MSSM} with large Higgs \ac{VEV} ratio $\tan\beta$,
typically their impact is limited (see e.g.\ \cite{Criado:2018thu} for a
comprehensive analysis). 

\begin{figure}
  \centering
  \includegraphics{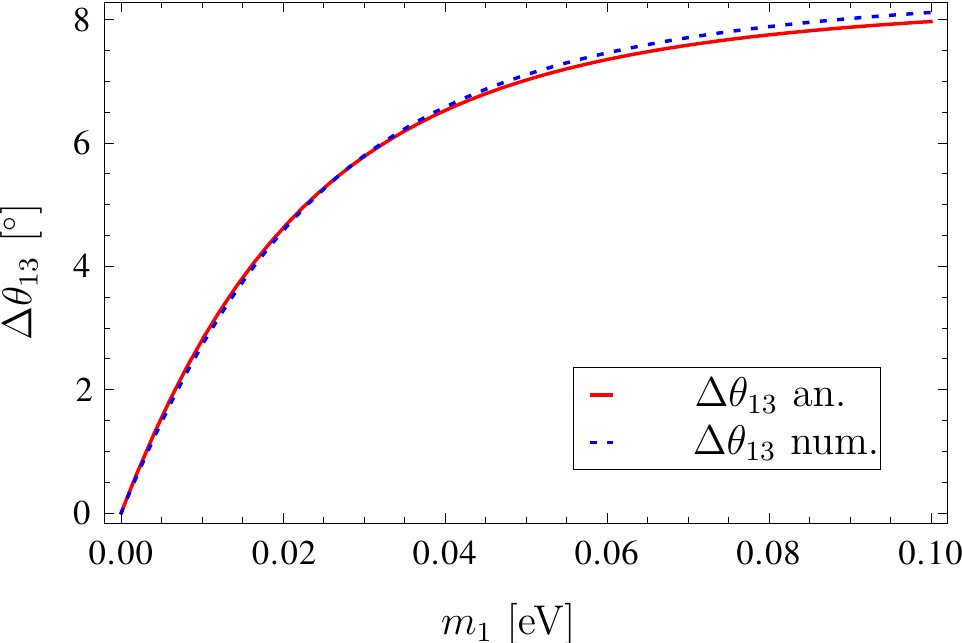}
  \caption{Change of $\theta_{13}$ due to the K\"ahler correction coming from
   the term $(L \otimes \AfourFlavonA)_{\rep{3}_\mathrm{a}}^\dagger\otimes 
   (L \otimes
   \AfourFlavonA)_{\rep{3}_\mathrm{s}}+\text{h.c.}$.  Here $m_1$ is  the
   smallest neutrino mass in \ac{NO}, the coefficient of this term has been 
   taken to be 1, and the ratio of \ac{VEV} over cutoff is $0.2$. The
   continuous line shows the result when using a linear approximation, while the
   dashed line shows the result of a numerical computation.  See Equation (3.8)
   and Figure 2 of \cite{Chen:2012ha}. Similar plots are obtained for the
   other mixing angles.}
  \label{fig:theta13}
\end{figure}

On the other hand, it is known that, in many models, corrections to the kinetic
terms have a significant impact on the predicted values of observables
\cite{Leurer:1993gy,Dudas:1995yu}. The main problem is that, in a pure
bottom--up \ac{EFT} approach, one cannot forbid higher--order terms in the 
K\"ahler potential that couple the flavons to the matter fields, nor the analogous 
terms in non--supersymmetric models. Those terms can induce the
$\mathcal{O}(\varepsilon)$ corrections which one expects to find from the above
qualitative argument. Specifically, for the $A_4$ model it has been shown that
the K\"ahler corrections are often far larger than the experimental
uncertainties in the mixing parameters~\cite{Chen:2012ha,Chen:2013aya}. In
particular, there are additional free, i.e.\ not predicted by the model,
parameters that allow one to adjust the predictions at will. 
We show an example in \Cref{fig:theta13}. As one can see, even under rather
conservative assumptions the corrections to the prediction exceed the
experimental error bars by far. In particular, including these terms, as one
should, can render the $A_4$ model from completely ruled out to perfectly
consistent. Likewise, these terms, which cannot be controlled in the bottom--up
approach, can also change consistent models to ruled--out constructions.

\subsubsection{\ac{VEV} alignment}
\label{sec:VEV_alignment}

Another subtle aspect of the model is the \ac{VEV} alignment. That is, rather
than postulating the \acp{VEV} \eqref{eq:VEValignmentA4}, there should be a
dynamical reason why they assume these values. It turns out that one can
sometimes construct models in which the desired \acp{VEV} emerge by minimizing a
flavon potential, and this is more straightforward if those \acp{VEV} respect
certain residual symmetries, as is the case in the $A_4$ model
\cite{Altarelli:2005yp,Bazzocchi:2007na,Feruglio:2009iu,King:2011zj,Holthausen:2011vd}. 
However, in practice, this often comes at the
expense of introducing ad hoc symmetries and extra fields. Some of these extra
fields are frequently referred to as ``driving fields'' but they really should
be regarded as Lagrange multipliers used to impose additional conditions on the
\acp{VEV} that sometimes appear to be ad hoc, i.e.\ have no other purpose than
creating the desired \acp{VEV}. Furthermore, these driving fields can introduce
more parameters to the theory which, in turn, reduce the predictive power
of the theory. Adding these driving fields introduces new ``free'' parameters,
which may or may not have a direct impact on the relevant predictions. Note also
that the corrections described in \Cref{sec:TraditionalCorrections} can also
affect the \ac{VEV} alignment. It has also been shown in~\cite{Kobayashi:2008ih}
that one may also achieve \ac{VEV} alignment by imposing boundary conditions of
scalar fields in extra dimensions. 

\subsection{\CP violation from finite groups}
\label{sec:CPfromFiniteGroups}

Let us briefly comment on another curious property of finite groups. It turns
out that finite groups may, unlike continuous (or Lie) groups, clash with
\CP~\cite{Chen:2009gf,Chen:2014tpa}. This means that some groups do not comply
with a physical \CP transformation. Notice that there is sometimes some
confusion in the literature. \CP transformations can be
generalized~\cite{Feruglio:2012cw,Holthausen:2012dk}. However, some of the
transformations that have been dubbed ``generalized \CP transformations'' do not
warrant \CP conservation, and may be more appropriately be referred to as
\CP--like transformations~\cite{Chen:2014tpa,Trautner:2016ezn}. On the other
hand, some groups do not admit proper \CP transformations~\cite{Chen:2014tpa},
and thus clash with \CP. It turns out that these \CP--violating groups are not
at all exotic, for instance, all odd--order finite groups are of this type. Yet
there are also even order groups of this type, such as $\Delta(54)$, which is
the traditional flavor symmetry~\cite{Kobayashi:2006wq} of some of the earliest
string models~\cite{Ibanez:1987sn}, where \CP violation is tied to the presence
of winding modes~\cite{Nilles:2018wex}, i.e.\ the very modes that are
instrumental for the \ac{UV} completion of the model. This leaves us with the
remarkable picture that flavor symmetries can very well be the reason why \CP is
violated, which also fits nicely to the observation that all of \CP violation in
Nature found so far resides in the flavor sector. One may thus hope to obtain
new solutions to the strong \CP problem~\cite{Ratz:2016scn}. However, so far a
concrete realization of this idea has been hindered by the limitations discussed
in \Cref{sec:TraditionalCorrections}. Yet the new ideas which we shall review in
\Cref{sec:ModularFlavorSymmetries} may provide us with the mileage required to
construct a concrete model.

\subsection{Origin of flavor symmetries}

While one can obviously impose flavor symmetries in a bottom--up approach, one
may wonder if they have a top--down motivation. The answer to this question is
affirmative. It has been shown that such symmetries can emerge in various string
models, such as heterotic orbifolds~\cite{Kobayashi:2006wq}, intersecting
D--branes~\cite{Abe:2009vi} and F--theory~\cite{Cvetic:2018xaq}. 

\subsection{Where to go from here?}

As we have seen, non--Abelian discrete flavor symmetries may motivate the
repetition of families, and provide us with a interpretation of the observed
mixing parameters. However, as discussed above, there are also limitations. One
major obstacle is \ac{VEV} alignment. As we shall see in the following section,
modular flavor symmetries evade some of the complications, and have an arguably
more direct connection to \ac{UV} completions of the \ac{SM}.

\section{Modular flavor symmetries}
\label{sec:ModularFlavorSymmetries}

A few years ago, Feruglio put forward a rather minimal and very successful model
\cite{Feruglio:2017spp} based on a modular version of the $A_4$ discussed in
\Cref{sec:ExampleA4}. This model has received significant attention  (cf.\
e.g.\ 
\cite{% 
Kobayashi:2018vbk,           %1803.10391 S3 and A4                lepton flavor
Criado:2018thu,              %1807.01125
deAnda:2018ecu,              %1812.05620 A4                       SU5 GUT in 6D T2/Z2 orbifold,
Okada:2018yrn,               %1812.09677 A4                       quark sector with CP violation
Ding:2019xna,                %1903.12588 A5                       neutrinos and charged leptons, Weinberg op./type I seesaw
Novichkov:2019sqv,           %1905.11970                          generalized CP and finite modular groups
Liu:2019khw,                 %1907.01488 Gamma'_N                 double cover, T' lepton masses and mixings
Kobayashi:2019xvz,           %1909.05139 S4 -> A4 by anomaly      modulus stabilization and lepton flavor
Asaka:2019vev,               %1909.06520 A4                       leptogenesis
Gui-JunDing:2019wap,         %1910.03460 S4 and A4                fixed points of Gamma_N, lepton flavor 
Kobayashi:2019uyt,           %1910.11553 A4                       modulus stabilization, CP violation  
Ding:2020yen,                %2003.13448 A4                       neutrino fitting with non-standard neutrino interactions
Liu:2020msy,                 %2007.13706 Metaplectic              neutrino fitting with half-integral modular weights
Ding:2020zxw,                            %Metaplectic
Yao:2020zml,
Novichkov:2021cgl
}) because it largely avoids the complications of \ac{VEV} alignment, and is, at
some level, able to make a large number of nontrivial predictions. This type of
models use the so--called modular flavor symmetries, which we shall review in
what follows.

\subsection{Modular transformations}

Let us first recall what modular transformations are. They can be thought of
as transformations that map a given torus on an equivalent torus. A 2--torus
$\mathds{T}^2$ emerges as the quotient $\mathds{C}/\mathds{Z}^2$. That is, two
points in the complex plane are identified if they differ by a lattice vector
$n_1\,e_1+n_2\,e_2$ with the $e_\alpha\in\mathds{C}$ denoting linearly
independent basis vectors and the $n_\alpha\in\mathds{Z}$ for
$\alpha\in\{1,2\}$. However, one can change the basis vectors without changing
the lattice. The allowed transformations are (cf.\ e.g.\ \cite{Kikuchi:2021ogn})
\begin{equation}\label{eq:TrafoBasisVectors} 
 \begin{pmatrix}
  e_2\\ e_1
 \end{pmatrix}
 \xmapsto{~\gamma~}
 \begin{pmatrix}
  e_2'\\ e_1'
 \end{pmatrix}
 =
 \begin{pmatrix}
  a & b \\ c & d
 \end{pmatrix}
 \begin{pmatrix}
  e_2\\ e_1
 \end{pmatrix}
 =:\gamma\,
  \begin{pmatrix}
  e_2\\ e_1
 \end{pmatrix}
 \;.
\end{equation}
Here, $\gamma$ is a $\text{SL}(2,\mathds{Z})$ matrix, i.e.\
\begin{equation}
 a,b,c,d\in\mathds{Z}\quad\text{and}\quad
 a\,d-b\,c=1\;.
\end{equation}
As is well known, the shape of a given torus is parametrized by the so--called
half period ratio $\tau:=e_2/e_1$. Without loss of generality one can demand
that $\im\tau>0$. Furthermore, under \eqref{eq:TrafoBasisVectors} $\tau$
undergoes the modular transformations
\begin{equation}\label{eq:ModularTrafoTau}
 \tau\xmapsto{~\gamma~} \frac{a\,\tau+b}{c\,\tau+d}\;.
\end{equation}

\subsection{Modular forms}

Holomorphic functions of $\tau$ complying with \eqref{eq:ModularTrafoTau} are
known to be rather constrained. We will be specifically interested in so--called
modular forms, which transform under $\gamma$ (cf.\ \Cref{eq:ModularTrafoTau})
as
\begin{equation}\label{eq:ModularForm}
 f(\tau)\xmapsto{~\gamma~} f\bigl(\gamma(\tau)\bigr)
 =f\left(\frac{a\,\tau+b}{c\,\tau+d}\right)
 =(c\,\tau+d)^k\,f(\tau)\;.
\end{equation}
Here, $k$ is the so--called modular weight, which is sometimes taken to be an
integer, but in top--down models often happens to be a rational
number.\footnote{For noninteger $k$ we are technically no longer dealing with
modular transformations, a point that we will get back to in
\Cref{sec:Metaplectic}.} In order to understand how this story is related to
finite groups, the perhaps most direct way is to consider the theory of 
vector--valued modular forms \cite{Liu:2021gwa}. The latter transform under
\eqref{eq:ModularTrafoTau} as
\begin{equation}\label{eq:VectorModularForm}
 Y(\tau)\xmapsto{~\gamma~} Y\bigl(\gamma(\tau)\bigr)
 =Y\left(\frac{a\,\tau+b}{c\,\tau+d}\right)
 =(c\,\tau+d)^k\,\rho(\gamma)\,Y(\tau)\;.
\end{equation} 
Here, $Y(\tau)=\bigl(Y_1(\tau),\dots,Y_d(\tau)\bigr)$ is a \rep{d}--plet, and
$\rho(\gamma)$ is a representation matrix of a finite modular group that arises
from the quotient of $\text{SL}(2,\mathds{Z})$ (or one of its multiple  covers)
divided by any of its normal subgroups.

\subsection{Modular flavor symmetries in the bottom-up approach}
\label{sec:MFSbottom-up}

In order to see how the modular flavor symmetries allow us to avoid  the
subtleties of \ac{VEV} alignment to large degree, let us review some aspects of
the Feruglio model~\cite{Feruglio:2017spp}. Rather than a ``traditional'' $A_4$
as in \Cref{sec:ExampleA4}, this model is based on a ``modular'' $A_4$. This
allows us to replace the $\AfourFlavonA$ flavon by a triplet of known functions
of $\tau$,
\begin{equation}
 \AfourFlavonA=\begin{pmatrix}
  (\AfourFlavonA)_1\\ (\AfourFlavonA)_2\\ (\AfourFlavonA)_3
 \end{pmatrix}
 \longrightarrow
 \begin{pmatrix}
  Y_1(\tau)\\ Y_2(\tau)\\ Y_3(\tau)
 \end{pmatrix}\;.
\end{equation}
That it, the crucial feature pointed out in \cite{Feruglio:2017spp} is that the
three complex components of the flavon $\AfourFlavonA$ can be replaced by three
functions $Y_i(\tau)$, which turn out to be modular forms building a  triplet
and transforming according to~\Cref{eq:VectorModularForm}, where  $\rho(\gamma)$
is the triplet representation of the finite modular symmetry $A_4\cong\Gamma_3$.
Importantly, these functions $Y_i(\tau)$ can be explicitly constructed.
Consequently, rather than aligning multiple \acp{VEV} comprising 6 real degrees
of freedom, one now faces the much more manageable challenge of fixing $\tau$,
which has two real degrees of freedom. As is well known for quite some time, for
largish $\im\tau$ these couplings are exponentially suppressed
\cite{Dine:1986zy,Dine:1987bq,Cvetic:1987qx}, which is why they appear at first
sight more suitable to accommodate the Yukawa couplings of the quarks and
charged leptons. However, Feruglio's fit~\cite{Feruglio:2017spp} wants $\tau$ to
be close to the so--called self--dual point $\tau = \ii$.  Fixing $\tau$ is part
of what is called moduli stabilization in string phenomenology, see
e.g.~\cite{Font:1990nt,Nilles:1990jv} for early references on this topic.
Remarkably, in these examples $\tau$ gets fixed at or close to the self--dual
point, i.e.\ where neutrino data wants it to be. See also
\cite{Novichkov:2022wvg} for a discussion of moduli fixing  and
\cite{Okada:2020ukr,Feruglio:2021dte,Novichkov:2021evw} for an analysis of
hierarchies around $\tau=\ii$ in the context of bottom--up modular flavor
symmetries. In this modular model there is no need for the $\xi$ flavon which
is instrumental in the model discussed in \Cref{sec:ExplicitA4Model} based
on a traditional $A_4$ symmetry.  Assuming diagonal charged lepton Yukawa
couplings, this model then derives 9 predictions from 3 input parameters:
\[ \begin{array}{rcl}
 \left.\begin{array}{@{}r@{}}
   \text{seesaw scale $\Lambda_\nu$}\\
   \re\tau\\
   \im\tau
 \end{array}\right\}
 & \longrightarrow &
 \left\{\begin{array}{@{}l@{}}
   \text{3 neutrino masses $m_i$}\\
   \text{3 mixing angles $\theta_{ij}$}\\
   \text{2 Majorana phases $\varphi_i$}\\
   \text{1 \CP phase $\delta_{\CP}$}
 \end{array}\right.
\end{array}\]
This system is overconstrained as we already know 5 observables, the two mass
squared differences and 3 mixing angles. It turns out that this model can
nonetheless fit observation surprisingly well. Furthermore, when introducing an
extra parameter, called $\varphi_3$, which changes the charged lepton Yukawa
matrix to a non--diagonal one, it is possible to get fit with $\Delta\chi^2 =
0.4$ for inverted ordering and $\Delta\chi^2 = 9.9 $ for normal ordering (see
Tables 4 and 5 of \cite{Criado:2018thu}, respectively). Here, $\Delta\chi^2$ is
obtained by comparing the model predictions to the current best--fit values of
the mixing angles, charged lepton Yukawa couplings and neutrino mass
differences.  \cite{Criado:2018thu} also find that the corrections from 
\ac{SUSY} breaking and \ac{RGE} corrections are relatively small in their model.
This model then makes testable predictions on the \CP phases as well as the
neutrino mass scale. This is a remarkable result.

It turns out that, similarly to what we already discussed in
\Cref{sec:TraditionalCorrections}, there are extra terms in this model which are
not fixed by the symmetries, and thus introduce additional free
parameters~\cite{Chen:2019ewa}. In some ways, this problem is even worse than in
the ``traditional'' case discussed in \Cref{sec:TraditionalCorrections} since
modular flavor symmetries are nonlinearly realized and thus provide us with less
expansion control. 

\begin{wrapfigure}[15]{r}[10pt]{7cm}
 \centering\vspace*{-1em}\includegraphics{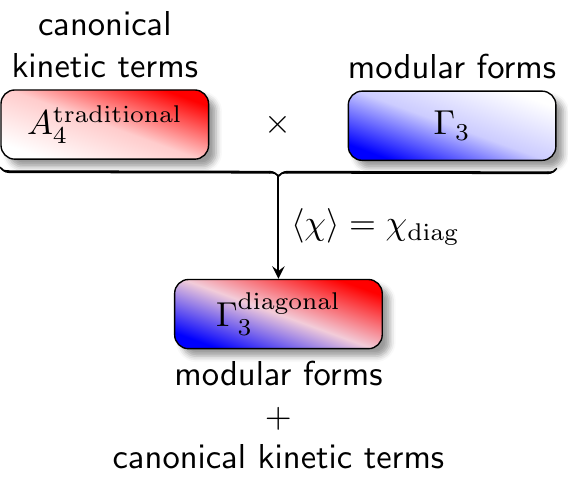}%
 \caption{``Quasi--eclectic''
  symmetries. Taken from \cite{Chen:2021prl}.}
 \label{fig:QuasiEclecticDiagram} 
\end{wrapfigure}
One can nonetheless show that, by borrowing inputs from the ``eclectic'' scheme
which we will discuss in \Cref{sec:Eclectic}, one can limit the impact of these
extra terms to be smaller than the current experimental uncertainties of the
observables \cite{Chen:2021prl}. In this variation of the Feruglio
model~\cite{Feruglio:2017spp}, a non--modular finite symmetry is  added, and the
full flavor symmetry is a product of a modular and a traditional symmetry. This
product is then broken to a diagonal subgroup (cf.\
\Cref{fig:QuasiEclecticDiagram}) which coincides with the modular $A_4$ of the
Feruglio model~\cite{Feruglio:2017spp}. There are still corrections of the order
$\Braket{\chi}/\Lambda$, where $\chi$ denotes the flavon accomplishing the
diagonal breaking, but this ratio can be as small as the $\tau$ Yukawa coupling,
which is of the order $10^{-2}$ unless $\tan\beta$ is large. Therefore, the
uncertainties of the predictions can be made comparable to the experimental
error bars. While this proof--of--principle example has been carefully crafted
to obtain sufficient control over the kinetic terms, the fact that it utilizes
ingredients first discussed in the top--down approach may be taken as an
indication that a combination of both approaches will ultimately provide us with
constructions which are both elegant, predictive and realistic.

\subsection{Metaplectic flavor symmetries}
\label{sec:Metaplectic}

\begin{figure}[t!]
 \centering
 \includegraphics[width=0.8\textwidth]{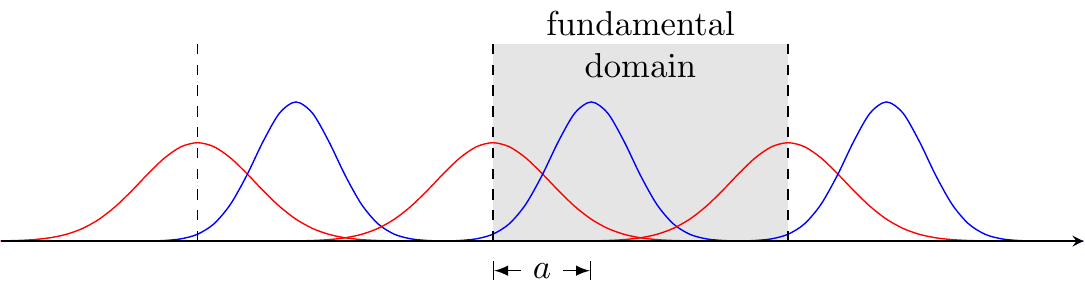}
 \caption{Overlap of two wavefunctions with separation $a$ on a torus. The
  overlap of a given, say red, curve is not just the overlap with one blue curve
  but with infinitely many of them due to the periodic nature of the torus. A
  representative space of the torus lattice is called the fundamental domain.
  Taken from \cite{Almumin:2021fbk}.}
 \label{fig:overlap}
\end{figure}

As we have argued in the introduction, the modular flavor symmetries are of
particular interest as they hint at top--down physics. A bit more practically,
one may want to have an interpretation, say, of the modular weight $k$ in
\Cref{eq:ModularForm,eq:VectorModularForm}, as well as of the origin of the
flavor symmetries. Magnetized tori \cite{Cremades:2004wa}, which are dual to
certain D--brane models \cite{Cremades:2003qj}, turn out to be an appropriate
playground to answer some of these questions
\cite{Kobayashi:2016ovu,Kobayashi:2018rad,Kobayashi:2018bff,Kariyazono:2019ehj,Ohki:2020bpo,Kikuchi:2020frp,Kikuchi:2020nxn,Kikuchi:2021ogn,Almumin:2021fbk,Tatsuta:2021deu,Kikuchi:2022bkn}.
In particular, the modular weights $k$ reflect the localization properties of
matter fields in the extra dimensions. 
This is because the K\"ahler metric, which depends on $\tau$ (as well as
other moduli) determines the normalization of the matter fields. Half--integer
modular weights then imply that the fields are something in between a bulk field
and a brane field. This is reflected by the profiles of the relevant zero
modes~\cite{Cremades:2004wa}. We depict the projection of some sample
wavefunctions in \Cref{fig:overlap}. This picture also offers an intuitive
understanding of the well--known exponential suppression of the couplings in the
large $\im\tau$ limit \cite{Dine:1986zy,Dine:1987bq,Cvetic:1987qx}.

Since modular weights are half--integers, the flavor symmetry is metaplectic
rather than modular~\cite{Almumin:2021fbk}. These metaplectic flavor symmetries
had been discussed in the bottom--up approach \cite{Yao:2020zml,Liu:2020msy},
and the expressions for the Yukawa couplings coincide with those obtained from
magnetized tori~\cite{Almumin:2021fbk}. That is, Yukawa couplings can be
computed in two ways, either in a bottom--up approach by postulating metaplectic
flavor symmetries and determining the corresponding modular forms, or in a
top--down way by computing the overlaps of wavefunctions on magnetized tori, and
the results agree. Although the successful models obtained in the bottom--up
approach \cite{Yao:2020zml,Liu:2020msy} can give predictions within the
experimental $1\sigma$ range (see e.g.\ Section 4.2 of \cite{Liu:2020msy}),
these have not yet been derived from the top--down approach. This is because one
cannot dial the data like modular weights and representations at will. Yet one
may take the phenomenologically successful constructions as a guide to choose
geometrical data of the top--down models to come closer to the real world. In
this sense, current and ongoing neutrino experiment data may provide us with
crucial insights in a possible \ac{UV} completion of the \ac{SM}.

\subsection{Eclectic flavor symmetries}
\label{sec:Eclectic}

Modular symmetries and modular forms appear quite naturally in top--down scenarios based 
on a class of string compactifications known as heterotic orbifolds~\cite{Lauer:1989ax,Lauer:1990tm}.
In these models, beyond the $3+1$ dimensions of our spacetime, the six extra dimensions of a string theory
assume the form of complex tori divided by discrete symmetries (see e.g.~\cite{Bailin:1999nk,Ramos-Sanchez:2008nwx,Vaudrevange:2008sm} 
for an introduction to these constructions). This procedure yields an \ac{UV} complete 
$3+1$--dimensional effective field theory endowed with various continuous and discrete global 
and gauge symmetries, which are fully determined by the shape of the compact dimensions. 
Furthermore, the matter spectrum together with their transformation properties under the 
available symmetries are also fixed by the compactification. 
In this kind of models, it has been shown that the exact matter spectrum of the \ac{SM} 
can be achieved, including quarks and leptons and their
mixings~\cite{Nilles:2008gq,Kappl:2010yu,Nilles:2014owa,Parr:2019bta}.

Among the symmetries of these string models, one identifies their discrete flavor symmetries (which 
technically correspond to the outer automorphisms of the Narain space group associated with the orbifold).
The origin and properties of these flavor symmetries have been explored in a series of papers
\cite{Nilles:2020gvu,Baur:2020jwc,Nilles:2020tdp,Nilles:2020kgo,Baur:2021mtl}.
The resulting symmetries lead to the so--called ``eclectic'' picture, which unifies the
traditional flavor symmetries of \Cref{sec:TraditionalFlavorSymmetries} with the
modular groups as in \Cref{sec:MFSbottom-up}. Roughly speaking, the eclectic flavor 
symmetry is given by
\begin{equation}
\label{eq:eclectic1}
 G_\text{eclectic}=G_\text{traditional}\cup G_\text{modular}\;,
\end{equation}
where ``$\cup$'' is to be understood as the multiplicative closure.
The eclectic symmetries include
\begin{itemize}
 \item traditional flavor symmetries,
 \item modular flavor symmetries,
 \item $R$ symmetries (including non--Abelian discrete $R$ symmetries), and
 \item $\CP$ symmetries and $\CP$--like transformations (see
  \Cref{sec:CPfromFiniteGroups} for the distinction).
\end{itemize}
Interestingly, both $R$ and $\CP$ (and $\CP$--like) symmetries are linked to modular 
transformations, arising from $\mathrm{SL}(2,\mathds{Z})$ or even 
$\mathrm{Sp}(2g,\mathds{Z})$~\cite{Baur:2021mtl}. (The latter has also been explored 
in the bottom--up approach~\cite{Ding:2020zxw}.) Further, all charges of matter fields, including
their modular weights $k\in\mathds Q$ and representations $\rho(\gamma)$ are fully fixed 
by the string construction. Whereas the modular weights of matter are fractional
in general, it turns out that the couplings among them are governed by 
integer positive modular weights, and are thus modular forms.

This eclectic picture provides a nontrivial mixture of symmetries, which
constrains not only the superpotential but also the K\"ahler potential; i.e.\
models with eclectic flavor symmetries are more restricted that bottom--up
constructions. In fact, these constraints solve the challenges on predictability
that bottom--up models endowed with flavor modular symmetries
face~\cite{Chen:2019ewa}. This advantage is partly due to the large number of
elements of eclectic flavor symmetries (e.g.\ the eclectic flavor group of a
$\mathds T^2/\Z2$ heterotic orbifold sector has 4608
elements~\cite{Baur:2021mtl}), but it actually follows from the natural 
appearance of a traditional flavor subgroup within eclectic symmetries. This
feature has been exploited e.g.\ in the bottom--up quasi--eclectic
scenario~\cite{Chen:2021prl} discussed in \Cref{sec:MFSbottom-up}.

Reproducing data requires the breakdown of the eclectic flavor symmetry. This is 
done in two steps: (1) $\tau$ is stabilized at a point in modular space, breaking 
the modular subgroup, and (2) one or more flavons develop VEVs, breaking the 
remaining traditional flavor symmetry. This leads to a rich
variety of flavor symmetry patterns~\cite{Baur:2021bly}, which can in some cases
match the mass and mixing textures of quarks and charged leptons observed in nature,
and yield predictions for the neutrino sector~\cite{Baur:2022toappear}, where all
corrections are under control.
Even though the flavon VEVs can, in principle, be fixed by demanding that supersymmetry 
is preserved at low energies in these string models, it is fair to say that finding a 
principle that dynamically fixes both the modulus and flavon VEVs at the right values for 
phenomenology is still an open question. Moreover, not all possible models based on 
eclectic flavor symmetries from strings have been phenomenologically investigated.
In anticipation of the upcoming neutrino data, it is important to pursue this
task now or in the near future.

One can generalize this top--down framework to arrive at bottom--up models endowed with
eclectic flavor symmetries. To achieve this goal one must stress two important
features of these symmetries in string constructions. First, it turns out 
that $G_\text{modular}$ is always a subgroup of the outer automorphisms of
$G_\text{traditional}$. Secondly, for a subgroup $G$ of $G_\text{traditional}$ 
\Cref{eq:eclectic1} takes the form
\begin{equation}
\label{eq:eclectic2}
 G_\text{eclectic}=G\rtimes G_\text{modular}\,,\qquad G\subset G_\text{traditional}\;,
\end{equation}
which means in particular that $G_\text{modular}$ and $G$ do not commute.
It has been found that these observations hold also in models based on magnetized
tori~\cite{Ohki:2020bpo}. Using these observations and the definitions of 
finite modular groups, a large class of bottom--up eclectic flavor symmetries
has been constructed~\cite{Nilles:2020nnc}. A relevant pending question is what
kind of neutrino phenomenology can be obtained from these symmetries. Further, it is 
clear now that more general eclectic flavor symmetries can be constructed, especially
including vector--valued modular forms~\cite{Liu:2021gwa}, which might open  
new avenues of relating neutrino data to possible \ac{UV} completions of the
\ac{SM}.

\subsection{Nonsupersymmetric modular flavor symmetries}

In our discussion so far we always assumed low--energy \ac{SUSY}, which,
however, we are far from certain that it is realized in Nature. It has been
argued in~\cite{Cremades:2004wa,Almumin:2021fbk} that low--energy \ac{SUSY} may,
in principle, not be required for the Yukawa  couplings to be of the metaplectic
form. Further, large classes of explicit string  models similar to those
presented in \Cref{sec:Eclectic} with the exact spectrum of the \ac{SM} and no
\ac{SUSY} have been
built~\cite{Blaszczyk:2014qoa,Ashfaque:2015vta,Abel:2015oxa,Perez-Martinez:2021zjj}. 
However, much more effort has to be devoted to better understand their details,
including the stability of these
models~\cite{Abel:2015oxa,GrootNibbelink:2017luf},  the versions of discrete
(traditional and modular) flavor  symmetries that they exhibit, and hence the
phenomenology they yield.

\section{Summary and Outlook}
\label{sec:Summary}

Our main summary is contained in the
\hyperlink{ExecutiveSummary}{Executive Summary}.
We have argued that, given the expected wealth of neutrino data, i.e.\
experimental knowledge of the flavor sector of the \ac{SM}, it is now the time
to sharpen our theoretical understanding of the matter. While measurements
become more precise, a fully convincing interpretation of the data has remained
elusive so far. The so--called modular flavor symmetries have a clear top--down
motivation and, at the same time, arguably provide us with some the most
compelling models of flavor. 
We have outlined how a successful unification of bottom--up and
top--down ideas and techniques may guide us towards a new era of precision flavor model
building in which future experimental results can give us crucial insights in
the \ac{UV} completion of the \ac{SM}.

\subsection*{Acknowledgments}

We would like to thank Yuri Shirman for discussions and Shirley Li for
correspondence. The work of Y.A.\ was supported by Kuwait University. The work
of M.C. is supported by US Department of Energy, under Grant No.\ DE-SC0015640.
The work of M.-C.C., V.K.-P., A.M., M.R.\ and S.S.\ is supported by the National
Science Foundation, under Grant No.\ PHY-1915005. This work is also supported by
UC-MEXUS-CONACyT grant No.\ CN-20-38. This work was performed in part at
Aspen Center for Physics, which is supported by National Science Foundation
grant PHY-1607611.

\begin{acronym}
  \acro{CnB}[C$\nu$B]{Cosmic Neutrino Background}
  \acro{EFT}{effective field theory}
  \acro{GUT}{Grand Unified Theory}
  \acro{IO}{inverted ordering}
  \acro{LHC}{Large Hadron Collider}
  \acro{MFV}{Minimal Flavor Violation}
  \acro{MSSM}{minimal supersymmetric standard model}
  \acro{NO}{normal ordering}
  \acro{NSI}{non--standard interactions}
  \acro{QFT}{quantum field theory}
  \acro{RGE}{renormalization group equation}
  \acro{SB}{symmetry based}
  \acro{SM}{Standard Model of Particle Physics}
  \acro{SUSY}{supersymmetry}
  \acro{TB}{torus based}
  \acro{UV}{ultraviolet}
  \acro{VEV}{vacuum expectation value}
\end{acronym}

\bibliography{./NFWP}

\providecommand{\href}[2]{#2}\begingroup\raggedright\begin{thebibliography}{100}

\bibitem{Ferrara:1989bc}
S.~Ferrara, D.~L{\"u}st, A.~D. Shapere, and S.~Theisen, ``{Modular Invariance
  in Supersymmetric Field Theories},''
  \href{http://dx.doi.org/10.1016/0370-2693(89)90583-2}{{\em Phys. Lett. B}
  {\bfseries 225} (1989) 363}.

\bibitem{Chun:1989se}
E.~J. Chun, J.~Mas, J.~Lauer, and H.~P. Nilles, ``{Duality and Landau-ginzburg
  Models},'' \href{http://dx.doi.org/10.1016/0370-2693(89)90630-8}{{\em Phys.
  Lett. B} {\bfseries 233} (1989) 141--146}.

\bibitem{Quevedo:1996sv}
F.~Quevedo, ``{Lectures on superstring phenomenology},''
  \href{http://dx.doi.org/10.1063/1.49735}{{\em AIP Conf. Proc.} {\bfseries
  359} (1996) 202--242}, \href{http://arxiv.org/abs/hep-th/9603074}{{\ttfamily
  arXiv:hep-th/9603074}}.

\bibitem{Feruglio:2017spp}
F.~Feruglio, \href{http://dx.doi.org/10.1142/9789813238053_0012}{``{Are
  neutrino masses modular forms?},''} in {\em From My Vast Repertoire ...:
  Guido Altarelli's Legacy}, A.~Levy, S.~Forte, and G.~Ridolfi, eds.,
  pp.~227--266.
\newblock 2019.
\newblock \href{http://arxiv.org/abs/1706.08749}{{\ttfamily arXiv:1706.08749
  [hep-ph]}}.
\newblock
{\texttt{https://arxiv.org/abs/1706.08749}}.
%%CITATION = ARXIV:1706.08749;%%.

\bibitem{Kaplan:1993ej}
D.~B. Kaplan and M.~Schmaltz, ``{Flavor unification and discrete nonAbelian
  symmetries},'' \href{http://dx.doi.org/10.1103/PhysRevD.49.3741}{{\em Phys.
  Rev. D} {\bfseries 49} (1994) 3741--3750},
  \href{http://arxiv.org/abs/hep-ph/9311281}{{\ttfamily arXiv:hep-ph/9311281}}.

\bibitem{Pontecorvo:1957qd}
B.~Pontecorvo, ``{Inverse beta processes and nonconservation of lepton
  charge},'' {\em Zh. Eksp. Teor. Fiz.} {\bfseries 34} (1957) 247.

\bibitem{Gribov:1968kq}
V.~N. Gribov and B.~Pontecorvo, ``{Neutrino astronomy and lepton charge},''
  \href{http://dx.doi.org/10.1016/0370-2693(69)90525-5}{{\em Phys. Lett. B}
  {\bfseries 28} (1969) 493}.

\bibitem{Maki:1962mu}
Z.~Maki, M.~Nakagawa, and S.~Sakata, ``{Remarks on the unified model of
  elementary particles},'' \href{http://dx.doi.org/10.1143/PTP.28.870}{{\em
  Prog. Theor. Phys.} {\bfseries 28} (1962) 870--880}.

\bibitem{Super-Kamiokande:1998kpq}
{\bfseries Super-Kamiokande} Collaboration, Y.~Fukuda {\em et al.}, ``{Evidence
  for oscillation of atmospheric neutrinos},''
  \href{http://dx.doi.org/10.1103/PhysRevLett.81.1562}{{\em Phys. Rev. Lett.}
  {\bfseries 81} (1998) 1562--1567},
  \href{http://arxiv.org/abs/hep-ex/9807003}{{\ttfamily arXiv:hep-ex/9807003}}.

\bibitem{SNO:2001kpb}
{\bfseries SNO} Collaboration, Q.~R. Ahmad {\em et al.}, ``{Measurement of the
  rate of $\nu_e+d \to p+p+e^-$ interactions produced by $^8$B solar neutrinos
  at the Sudbury Neutrino Observatory},''
  \href{http://dx.doi.org/10.1103/PhysRevLett.87.071301}{{\em Phys. Rev. Lett.}
  {\bfseries 87} (2001) 071301},
  \href{http://arxiv.org/abs/nucl-ex/0106015}{{\ttfamily
  arXiv:nucl-ex/0106015}}.

\bibitem{Esteban:2020cvm}
I.~Esteban, M.~C. Gonz{\'a}lez-Garc{\'i}a, M.~Maltoni, T.~Schwetz, and A.~Zhou,
  ``{The fate of hints: updated global analysis of three-flavor neutrino
  oscillations},'' \href{http://dx.doi.org/10.1007/JHEP09(2020)178}{{\em JHEP}
  {\bfseries 09} (2020) 178}, \href{http://arxiv.org/abs/2007.14792}{{\ttfamily
  arXiv:2007.14792 [hep-ph]}}.

\bibitem{NuFit:2021}
I.~Esteban, M.~C. Gonz{\'a}lez-Garc{\'i}a, M.~Maltoni, T.~Schwetz, and A.~Zhou.
\newblock \texttt{http://www.nu-fit.org/}.

\bibitem{T2K:2019bcf}
{\bfseries T2K} Collaboration, K.~Abe {\em et al.}, ``{Constraint on the
  matter\textendash{}antimatter symmetry-violating phase in neutrino
  oscillations},'' \href{http://dx.doi.org/10.1038/s41586-020-2177-0}{{\em
  Nature} {\bfseries 580} no.~7803, (2020) 339--344},
  \href{http://arxiv.org/abs/1910.03887}{{\ttfamily arXiv:1910.03887
  [hep-ex]}}. [Erratum: Nature 583, E16 (2020)].

\bibitem{Esteban:2018azc}
I.~Esteban, M.~C. Gonz{\'a}lez-Garc{\'i}a, A.~Hern{\'a}ndez-Cabezudo,
  M.~Maltoni, and T.~Schwetz, ``{Global analysis of three-flavour neutrino
  oscillations: synergies and tensions in the determination of $\theta_{23}$,
  $\delta_{CP}$, and the mass ordering},''
  \href{http://dx.doi.org/10.1007/JHEP01(2019)106}{{\em JHEP} {\bfseries 01}
  (2019) 106}, \href{http://arxiv.org/abs/1811.05487}{{\ttfamily
  arXiv:1811.05487 [hep-ph]}}.

\bibitem{Chatterjee:2022nia}
S.~S. Chatterjee and A.~Palazzo, ``{Resolving the NOvA and T2K tension in the
  presence of Neutrino Non-Standard Interactions},''
  \href{http://dx.doi.org/10.22323/1.402.0059}{{\em PoS} {\bfseries NuFact2021}
  (2022) 059}, \href{http://arxiv.org/abs/2201.10412}{{\ttfamily
  arXiv:2201.10412 [hep-ph]}}.

\bibitem{NOvA:2021nfi}
{\bfseries NOvA} Collaboration, M.~A. Acero {\em et al.}, ``{An Improved
  Measurement of Neutrino Oscillation Parameters by the NOvA Experiment},''
  \href{http://arxiv.org/abs/2108.08219}{{\ttfamily arXiv:2108.08219
  [hep-ex]}}.

\bibitem{Hyper-Kamiokande:2018ofw}
{\bfseries Hyper-Kamiokande} Collaboration, K.~Abe {\em et al.},
  ``{Hyper-Kamiokande Design Report},''
  \href{http://arxiv.org/abs/1805.04163}{{\ttfamily arXiv:1805.04163
  [physics.ins-det]}}.

\bibitem{DUNE:2016hlj}
{\bfseries DUNE} Collaboration, R.~Acciarri {\em et al.}, ``{Long-Baseline
  Neutrino Facility (LBNF) and Deep Underground Neutrino Experiment (DUNE)}:
  {Conceptual Design Report, Volume 1: The LBNF and DUNE Projects},''
  \href{http://arxiv.org/abs/1601.05471}{{\ttfamily arXiv:1601.05471
  [physics.ins-det]}}.

\bibitem{Smirnov:2018ywm}
M.~V. Smirnov, Z.~J. Hu, S.~J. Li, and J.~J. Ling, ``{The possibility of
  leptonic CP-violation measurement with JUNO},''
  \href{http://dx.doi.org/10.1016/j.nuclphysb.2018.05.003}{{\em Nucl. Phys. B}
  {\bfseries 931} (2018) 437--445},
  \href{http://arxiv.org/abs/1802.03677}{{\ttfamily arXiv:1802.03677
  [hep-ph]}}.

\bibitem{Song:2020nfh}
N.~Song, S.~W. Li, C.~A. Arg\"uelles, M.~Bustamante, and A.~C. Vincent, ``{The
  Future of High-Energy Astrophysical Neutrino Flavor Measurements},''
  \href{http://dx.doi.org/10.1088/1475-7516/2021/04/054}{{\em JCAP} {\bfseries
  04} (2021) 054}, \href{http://arxiv.org/abs/2012.12893}{{\ttfamily
  arXiv:2012.12893 [hep-ph]}}.

\bibitem{Denton:2019ovn}
P.~B. Denton, S.~J. Parke, and X.~Zhang, ``{Neutrino oscillations in matter via
  eigenvalues},'' \href{http://dx.doi.org/10.1103/PhysRevD.101.093001}{{\em
  Phys. Rev. D} {\bfseries 101} no.~9, (2020) 093001},
  \href{http://arxiv.org/abs/1907.02534}{{\ttfamily arXiv:1907.02534
  [hep-ph]}}.

\bibitem{Qian:2015waa}
X.~Qian and P.~Vogel, ``{Neutrino Mass Hierarchy},''
  \href{http://dx.doi.org/10.1016/j.ppnp.2015.05.002}{{\em Prog. Part. Nucl.
  Phys.} {\bfseries 83} (2015) 1--30},
  \href{http://arxiv.org/abs/1505.01891}{{\ttfamily arXiv:1505.01891
  [hep-ex]}}.

\bibitem{Lee_2016}
C.~M. Lee and J.~H. Selby, ``Higher-Order Interference in Extensions of Quantum
  Theory,'' \href{http://dx.doi.org/10.1007/s10701-016-0045-4}{{\em Foundations
  of Physics} {\bfseries 47} no.~1, (Oct, 2016) 89--112}.
  \url{https://doi.org/10.1007\%2Fs10701-016-0045-4}.

\bibitem{Xu:2020pzr}
B.~Xu, ``{Neutrino Decoherence in Simple Open Quantum Systems},''
  \href{http://arxiv.org/abs/2009.13471}{{\ttfamily arXiv:2009.13471
  [hep-ph]}}.

\bibitem{Agostini:2022zub}
M.~Agostini, G.~Benato, J.~A. Detwiler, J.~Men\'endez, and F.~Vissani,
  ``{Toward the discovery of matter creation with neutrinoless double-beta
  decay},'' \href{http://arxiv.org/abs/2202.01787}{{\ttfamily arXiv:2202.01787
  [hep-ex]}}.

\bibitem{Cirigliano:2022oqy}
V.~Cirigliano {\em et al.}, ``{Neutrinoless Double-Beta Decay: A Roadmap for
  Matching Theory to Experiment},''
  \href{http://arxiv.org/abs/2203.12169}{{\ttfamily arXiv:2203.12169
  [hep-ph]}}.

\bibitem{Gastaldo:2017edk}
L.~Gastaldo {\em et al.}, ``{The electron capture in$^{163}$Ho experiment
  \textendash{} ECHo},''
  \href{http://dx.doi.org/10.1140/epjst/e2017-70071-y}{{\em Eur. Phys. J. ST}
  {\bfseries 226} no.~8, (2017) 1623--1694}.

\bibitem{PTOLEMY:2019hkd}
{\bfseries PTOLEMY} Collaboration, M.~G. Betti {\em et al.}, ``{Neutrino
  physics with the PTOLEMY project: active neutrino properties and the light
  sterile case},'' \href{http://dx.doi.org/10.1088/1475-7516/2019/07/047}{{\em
  JCAP} {\bfseries 07} (2019) 047},
  \href{http://arxiv.org/abs/1902.05508}{{\ttfamily arXiv:1902.05508
  [astro-ph.CO]}}.

\bibitem{Project8:2017nal}
{\bfseries Project 8} Collaboration, A.~Ashtari~Esfahani {\em et al.},
  ``{Determining the neutrino mass with cyclotron radiation emission
  spectroscopy\textemdash{}Project 8},''
  \href{http://dx.doi.org/10.1088/1361-6471/aa5b4f}{{\em J. Phys. G} {\bfseries
  44} no.~5, (2017) 054004}, \href{http://arxiv.org/abs/1703.02037}{{\ttfamily
  arXiv:1703.02037 [physics.ins-det]}}.

\bibitem{Danilov:2022str}
M.~Danilov, ``{Review of sterile neutrino searches at very short-baseline
  reactor experiments},'' \href{http://arxiv.org/abs/2203.03042}{{\ttfamily
  arXiv:2203.03042 [hep-ex]}}.

\bibitem{Coloma:2022avw}
P.~Coloma, I.~Esteban, M.~C. Gonz{\'a}lez-Garc{\'i}a, L.~Larizgoitia,
  F.~Monrabal, and S.~Palomares-Ruiz, ``{Bounds on new physics with data of the
  Dresden-II reactor experiment and COHERENT},''
  \href{http://arxiv.org/abs/2202.10829}{{\ttfamily arXiv:2202.10829
  [hep-ph]}}.

\bibitem{Minkowski:1977sc}
P.~Minkowski, ``{$\mu \to e\gamma$ at a Rate of One Out of $10^{9}$ Muon
  Decays?},'' \href{http://dx.doi.org/10.1016/0370-2693(77)90435-X}{{\em Phys.
  Lett. B} {\bfseries 67} (1977) 421--428}.

\bibitem{Yanagida:1979as}
T.~Yanagida, ``{Horizontal gauge symmetry and masses of neutrinos},'' {\em
  Conf. Proc. C} {\bfseries 7902131} (1979) 95--99.

\bibitem{Glashow:1979nm}
S.~L. Glashow, ``{The Future of Elementary Particle Physics},''
  \href{http://dx.doi.org/10.1007/978-1-4684-7197-7_15}{{\em NATO Sci. Ser. B}
  {\bfseries 61} (1980) 687}.

\bibitem{Gell-Mann:1979vob}
M.~Gell-Mann, P.~Ramond, and R.~Slansky, ``{Complex Spinors and Unified
  Theories},'' {\em Conf. Proc. C} {\bfseries 790927} (1979) 315--321,
  \href{http://arxiv.org/abs/1306.4669}{{\ttfamily arXiv:1306.4669 [hep-th]}}.

\bibitem{Magg:1980ut}
M.~Magg and C.~Wetterich, ``{Neutrino Mass Problem and Gauge Hierarchy},''
  \href{http://dx.doi.org/10.1016/0370-2693(80)90825-4}{{\em Phys. Lett. B}
  {\bfseries 94} (1980) 61--64}.

\bibitem{Lazarides:1980rn}
G.~Lazarides and Q.~Shafi, ``{Neutrino Masses in SU(5)},''
  \href{http://dx.doi.org/10.1016/0370-2693(81)90962-X}{{\em Phys. Lett. B}
  {\bfseries 99} (1981) 113--116}.

\bibitem{Mohapatra:1979ia}
R.~N. Mohapatra and G.~Senjanovic, ``{Neutrino Mass and Spontaneous Parity
  Violation},''
\href{http://dx.doi.org/10.1103/PhysRevLett.44.912}{{\em Phys. Rev. Lett.}
  {\bfseries 44} (1980) 912}.
%%CITATION = PRLTA,44,912;%%.

\bibitem{Mohapatra:1980yp}
R.~N. Mohapatra and G.~Senjanovic, ``{Neutrino Masses and Mixings in Gauge
  Models with Spontaneous Parity Violation},''
  \href{http://dx.doi.org/10.1103/PhysRevD.23.165}{{\em Phys. Rev. D}
  {\bfseries 23} (1981) 165}.

\bibitem{Foot:1988aq}
R.~Foot, H.~Lew, X.~G. He, and G.~C. Joshi, ``{Seesaw Neutrino Masses Induced
  by a Triplet of Leptons},'' \href{http://dx.doi.org/10.1007/BF01415558}{{\em
  Z. Phys. C} {\bfseries 44} (1989) 441}.

\bibitem{Fritzsch:1974nn}
H.~Fritzsch and P.~Minkowski, ``{Unified Interactions of Leptons and
  Hadrons},'' \href{http://dx.doi.org/10.1016/0003-4916(75)90211-0}{{\em Annals
  Phys.} {\bfseries 93} (1975) 193--266}.

\bibitem{Cai:2017jrq}
Y.~Cai, J.~Herrero-Garc\'\i{}a, M.~A. Schmidt, A.~Vicente, and R.~R. Volkas,
  ``{From the trees to the forest: a review of radiative neutrino mass
  models},'' \href{http://dx.doi.org/10.3389/fphy.2017.00063}{{\em Front. in
  Phys.} {\bfseries 5} (2017) 63},
  \href{http://arxiv.org/abs/1706.08524}{{\ttfamily arXiv:1706.08524
  [hep-ph]}}.

\bibitem{Zee:1980ai}
A.~Zee, ``{A Theory of Lepton Number Violation, Neutrino Majorana Mass, and
  Oscillation},'' \href{http://dx.doi.org/10.1016/0370-2693(80)90349-4}{{\em
  Phys. Lett. B} {\bfseries 93} (1980) 389}. [Erratum: Phys.Lett.B 95, 461
  (1980)].

\bibitem{Ma:2006km}
E.~Ma, ``{Verifiable radiative seesaw mechanism of neutrino mass and dark
  matter},'' \href{http://dx.doi.org/10.1103/PhysRevD.73.077301}{{\em Phys.
  Rev. D} {\bfseries 73} (2006) 077301},
  \href{http://arxiv.org/abs/hep-ph/0601225}{{\ttfamily arXiv:hep-ph/0601225}}.

\bibitem{Arkani-Hamed:2000oup}
N.~Arkani-Hamed, L.~J. Hall, H.~Murayama, D.~Tucker-Smith, and N.~Weiner,
  ``{Small neutrino masses from supersymmetry breaking},''
  \href{http://dx.doi.org/10.1103/PhysRevD.64.115011}{{\em Phys. Rev. D}
  {\bfseries 64} (2001) 115011},
  \href{http://arxiv.org/abs/hep-ph/0006312}{{\ttfamily arXiv:hep-ph/0006312}}.

\bibitem{Babu:1988yq}
K.~S. Babu and X.~G. He, ``{DIRAC NEUTRINO MASSES AS TWO LOOP RADIATIVE
  CORRECTIONS},'' \href{http://dx.doi.org/10.1142/S0217732389000095}{{\em Mod.
  Phys. Lett. A} {\bfseries 4} (1989) 61}.

\bibitem{Farzan:2012ev}
Y.~Farzan, S.~Pascoli, and M.~A. Schmidt, ``{Recipes and Ingredients for
  Neutrino Mass at Loop Level},''
  \href{http://dx.doi.org/10.1007/JHEP03(2013)107}{{\em JHEP} {\bfseries 03}
  (2013) 107}, \href{http://arxiv.org/abs/1208.2732}{{\ttfamily arXiv:1208.2732
  [hep-ph]}}.

\bibitem{Grossman:1999ra}
Y.~Grossman and M.~Neubert, ``{Neutrino masses and mixings in nonfactorizable
  geometry},'' \href{http://dx.doi.org/10.1016/S0370-2693(00)00054-X}{{\em
  Phys. Lett. B} {\bfseries 474} (2000) 361--371},
  \href{http://arxiv.org/abs/hep-ph/9912408}{{\ttfamily arXiv:hep-ph/9912408}}.

\bibitem{Huber:2000ie}
S.~J. Huber and Q.~Shafi, ``{Fermion masses, mixings and proton decay in a
  Randall-Sundrum model},''
  \href{http://dx.doi.org/10.1016/S0370-2693(00)01399-X}{{\em Phys. Lett. B}
  {\bfseries 498} (2001) 256--262},
  \href{http://arxiv.org/abs/hep-ph/0010195}{{\ttfamily arXiv:hep-ph/0010195}}.

\bibitem{Park:2017yrn}
S.~C. Park and C.~S. Shin, ``{Clockwork seesaw mechanisms},''
  \href{http://dx.doi.org/10.1016/j.physletb.2017.11.057}{{\em Phys. Lett. B}
  {\bfseries 776} (2018) 222--226},
  \href{http://arxiv.org/abs/1707.07364}{{\ttfamily arXiv:1707.07364
  [hep-ph]}}.

\bibitem{Hong:2019bki}
S.~Hong, G.~Kurup, and M.~Perelstein, ``{Clockwork Neutrinos},''
  \href{http://dx.doi.org/10.1007/JHEP10(2019)073}{{\em JHEP} {\bfseries 10}
  (2019) 073}, \href{http://arxiv.org/abs/1903.06191}{{\ttfamily
  arXiv:1903.06191 [hep-ph]}}.

\bibitem{Babu:2020tnf}
K.~S. Babu and S.~Saad, ``{Flavor Hierarchies from Clockwork in SO(10) GUT},''
  \href{http://dx.doi.org/10.1103/PhysRevD.103.015009}{{\em Phys. Rev. D}
  {\bfseries 103} no.~1, (2021) 015009},
  \href{http://arxiv.org/abs/2007.16085}{{\ttfamily arXiv:2007.16085
  [hep-ph]}}.

\bibitem{Babu:2001ex}
K.~S. Babu and C.~N. Leung, ``{Classification of effective neutrino mass
  operators},'' \href{http://dx.doi.org/10.1016/S0550-3213(01)00504-1}{{\em
  Nucl. Phys. B} {\bfseries 619} (2001) 667--689},
  \href{http://arxiv.org/abs/hep-ph/0106054}{{\ttfamily arXiv:hep-ph/0106054}}.

\bibitem{Giedt:2005vx}
J.~Giedt, G.~L. Kane, P.~Langacker, and B.~D. Nelson, ``{Massive neutrinos and
  (heterotic) string theory},''
  \href{http://dx.doi.org/10.1103/PhysRevD.71.115013}{{\em Phys. Rev. D}
  {\bfseries 71} (2005) 115013},
  \href{http://arxiv.org/abs/hep-th/0502032}{{\ttfamily arXiv:hep-th/0502032}}.

\bibitem{Blumenhagen:2006xt}
R.~Blumenhagen, M.~Cveti{\v{c}}, and T.~Weigand, ``{Spacetime instanton
  corrections in 4D string vacua: The Seesaw mechanism for D-Brane models},''
  \href{http://dx.doi.org/10.1016/j.nuclphysb.2007.02.016}{{\em Nucl. Phys. B}
  {\bfseries 771} (2007) 113--142},
  \href{http://arxiv.org/abs/hep-th/0609191}{{\ttfamily arXiv:hep-th/0609191}}.

\bibitem{Buchmuller:2007zd}
W.~Buchm{\"u}ller, K.~Hamaguchi, O.~Lebedev, S.~Ramos-S{\'a}nchez, and M.~Ratz,
  ``{Seesaw neutrinos from the heterotic string},''
  \href{http://dx.doi.org/10.1103/PhysRevLett.99.021601}{{\em Phys. Rev. Lett.}
  {\bfseries 99} (2007) 021601},
  \href{http://arxiv.org/abs/hep-ph/0703078}{{\ttfamily arXiv:hep-ph/0703078}}.

\bibitem{Feldstein:2011ck}
B.~Feldstein and W.~Klemm, ``{Large Mixing Angles From Many Right-Handed
  Neutrinos},'' \href{http://dx.doi.org/10.1103/PhysRevD.85.053007}{{\em Phys.
  Rev. D} {\bfseries 85} (2012) 053007},
  \href{http://arxiv.org/abs/1111.6690}{{\ttfamily arXiv:1111.6690 [hep-ph]}}.

\bibitem{Hall:1999sn}
L.~J. Hall, H.~Murayama, and N.~Weiner, ``{Neutrino mass anarchy},''
  \href{http://dx.doi.org/10.1103/PhysRevLett.84.2572}{{\em Phys. Rev. Lett.}
  {\bfseries 84} (2000) 2572--2575},
  \href{http://arxiv.org/abs/hep-ph/9911341}{{\ttfamily arXiv:hep-ph/9911341}}.

\bibitem{deGouvea:2003xe}
A.~de~Gouvea and H.~Murayama, ``{Statistical test of anarchy},''
  \href{http://dx.doi.org/10.1016/j.physletb.2003.08.045}{{\em Phys. Lett. B}
  {\bfseries 573} (2003) 94--100},
  \href{http://arxiv.org/abs/hep-ph/0301050}{{\ttfamily arXiv:hep-ph/0301050}}.

\bibitem{deGouvea:2012ac}
A.~de~Gouvea and H.~Murayama, ``{Neutrino Mixing Anarchy: Alive and Kicking},''
  \href{http://dx.doi.org/10.1016/j.physletb.2015.06.028}{{\em Phys. Lett. B}
  {\bfseries 747} (2015) 479--483},
  \href{http://arxiv.org/abs/1204.1249}{{\ttfamily arXiv:1204.1249 [hep-ph]}}.

\bibitem{Ishimori:2010au}
H.~Ishimori, T.~Kobayashi, H.~Ohki, Y.~Shimizu, H.~Okada, and M.~Tanimoto,
  ``{Non-Abelian Discrete Symmetries in Particle Physics},''
  \href{http://dx.doi.org/10.1143/PTPS.183.1}{{\em Prog. Theor. Phys. Suppl.}
  {\bfseries 183} (2010) 1--163},
  \href{http://arxiv.org/abs/1003.3552}{{\ttfamily arXiv:1003.3552 [hep-th]}}.

\bibitem{Araki:2006sqx}
T.~Araki, ``{Anomaly of Discrete Symmetries and Gauge Coupling Unification},''
  \href{http://dx.doi.org/10.1143/PTP.117.1119}{{\em Prog. Theor. Phys.}
  {\bfseries 117} (2007) 1119--1138},
  \href{http://arxiv.org/abs/hep-ph/0612306}{{\ttfamily arXiv:hep-ph/0612306}}.

\bibitem{Araki:2008ek}
T.~Araki, T.~Kobayashi, J.~Kubo, S.~Ramos-S{\'a}nchez, M.~Ratz, and P.~K.~S.
  Vaudrevange, ``{(Non-)Abelian discrete anomalies},''
  \href{http://dx.doi.org/10.1016/j.nuclphysb.2008.07.005}{{\em Nucl. Phys. B}
  {\bfseries 805} (2008) 124--147},
  \href{http://arxiv.org/abs/0805.0207}{{\ttfamily arXiv:0805.0207 [hep-th]}}.

\bibitem{Chen:2015aba}
M.-C. Chen, M.~Fallbacher, M.~Ratz, A.~Trautner, and P.~K.~S. Vaudrevange,
  ``{Anomaly-safe discrete groups},''
  \href{http://dx.doi.org/10.1016/j.physletb.2015.05.047}{{\em Phys. Lett. B}
  {\bfseries 747} (2015) 22--26},
  \href{http://arxiv.org/abs/1504.03470}{{\ttfamily arXiv:1504.03470
  [hep-ph]}}.

\bibitem{Talbert:2018nkq}
J.~Talbert, ``{Pocket Formulae for Non-Abelian Discrete Anomaly Freedom},''
  \href{http://dx.doi.org/10.1016/j.physletb.2018.10.025}{{\em Phys. Lett. B}
  {\bfseries 786} (2018) 426--431},
  \href{http://arxiv.org/abs/1804.04237}{{\ttfamily arXiv:1804.04237
  [hep-ph]}}.

\bibitem{Kobayashi:2021xfs}
T.~Kobayashi and H.~Uchida, ``{Anomaly of non-Abelian discrete symmetries},''
  \href{http://dx.doi.org/10.1103/PhysRevD.105.036018}{{\em Phys. Rev. D}
  {\bfseries 105} no.~3, (2022) 036018},
  \href{http://arxiv.org/abs/2111.10811}{{\ttfamily arXiv:2111.10811
  [hep-th]}}.

\bibitem{Gripaios:2022vvc}
B.~Gripaios, ``{Gauge anomalies of finite groups},''
  \href{http://arxiv.org/abs/2201.11801}{{\ttfamily arXiv:2201.11801
  [hep-th]}}.

\bibitem{Csaki:1997aw}
C.~Cs{\'a}ki and H.~Murayama, ``{Discrete anomaly matching},''
  \href{http://dx.doi.org/10.1016/S0550-3213(97)00839-0}{{\em Nucl. Phys. B}
  {\bfseries 515} (1998) 114--162},
  \href{http://arxiv.org/abs/hep-th/9710105}{{\ttfamily arXiv:hep-th/9710105}}.

\bibitem{Henning:2021ctv}
B.~Henning, X.~Lu, T.~Melia, and H.~Murayama, ``{Outer automorphism
  anomalies},'' \href{http://dx.doi.org/10.1007/JHEP02(2022)094}{{\em JHEP}
  {\bfseries 02} (2022) 094}, \href{http://arxiv.org/abs/2111.04728}{{\ttfamily
  arXiv:2111.04728 [hep-th]}}.

\bibitem{Feruglio:2019ybq}
F.~Feruglio and A.~Romanino, ``{Lepton flavor symmetries},''
  \href{http://dx.doi.org/10.1103/RevModPhys.93.015007}{{\em Rev. Mod. Phys.}
  {\bfseries 93} no.~1, (2021) 015007},
  \href{http://arxiv.org/abs/1912.06028}{{\ttfamily arXiv:1912.06028
  [hep-ph]}}.

\bibitem{Ma:2001dn}
E.~Ma and G.~Rajasekaran, ``{Softly broken A(4) symmetry for nearly degenerate
  neutrino masses},'' \href{http://dx.doi.org/10.1103/PhysRevD.64.113012}{{\em
  Phys. Rev. D} {\bfseries 64} (2001) 113012},
  \href{http://arxiv.org/abs/hep-ph/0106291}{{\ttfamily arXiv:hep-ph/0106291}}.

\bibitem{Babu:2002dz}
K.~S. Babu, E.~Ma, and J.~W.~F. Valle, ``{Underlying A(4) symmetry for the
  neutrino mass matrix and the quark mixing matrix},''
  \href{http://dx.doi.org/10.1016/S0370-2693(02)03153-2}{{\em Phys. Lett. B}
  {\bfseries 552} (2003) 207--213},
  \href{http://arxiv.org/abs/hep-ph/0206292}{{\ttfamily arXiv:hep-ph/0206292}}.

\bibitem{Hirsch:2003dr}
M.~Hirsch, J.~C. Romao, S.~Skadhauge, J.~W.~F. Valle, and A.~Villanova~del
  Moral, ``{Phenomenological tests of supersymmetric A(4) family symmetry model
  of neutrino mass},'' \href{http://dx.doi.org/10.1103/PhysRevD.69.093006}{{\em
  Phys. Rev. D} {\bfseries 69} (2004) 093006},
  \href{http://arxiv.org/abs/hep-ph/0312265}{{\ttfamily arXiv:hep-ph/0312265}}.

\bibitem{Altarelli:2005yp}
G.~Altarelli and F.~Feruglio, ``{Tri-bimaximal neutrino mixing from discrete
  symmetry in extra dimensions},''
  \href{http://dx.doi.org/10.1016/j.nuclphysb.2005.05.005}{{\em Nucl. Phys. B}
  {\bfseries 720} (2005) 64--88},
  \href{http://arxiv.org/abs/hep-ph/0504165}{{\ttfamily arXiv:hep-ph/0504165}}.

\bibitem{Harrison:2002er}
P.~F. Harrison, D.~H. Perkins, and W.~G. Scott, ``{Tri-bimaximal mixing and the
  neutrino oscillation data},''
  \href{http://dx.doi.org/10.1016/S0370-2693(02)01336-9}{{\em Phys. Lett. B}
  {\bfseries 530} (2002) 167},
  \href{http://arxiv.org/abs/hep-ph/0202074}{{\ttfamily arXiv:hep-ph/0202074}}.

\bibitem{Antusch:2005gp}
S.~Antusch, J.~Kersten, M.~Lindner, M.~Ratz, and M.~A. Schmidt, ``{Running
  neutrino mass parameters in see-saw scenarios},''
  \href{http://dx.doi.org/10.1088/1126-6708/2005/03/024}{{\em JHEP} {\bfseries
  03} (2005) 024}, \href{http://arxiv.org/abs/hep-ph/0501272}{{\ttfamily
  arXiv:hep-ph/0501272}}.

\bibitem{Criado:2018thu}
J.~C. Criado and F.~Feruglio, ``{Modular Invariance Faces Precision Neutrino
  Data},'' \href{http://dx.doi.org/10.21468/SciPostPhys.5.5.042}{{\em SciPost
  Phys.} {\bfseries 5} no.~5, (2018) 042},
  \href{http://arxiv.org/abs/1807.01125}{{\ttfamily arXiv:1807.01125
  [hep-ph]}}.

\bibitem{Leurer:1993gy}
M.~Leurer, Y.~Nir, and N.~Seiberg, ``{Mass matrix models: The Sequel},''
  \href{http://dx.doi.org/10.1016/0550-3213(94)90074-4}{{\em Nucl. Phys. B}
  {\bfseries 420} (1994) 468--504},
  \href{http://arxiv.org/abs/hep-ph/9310320}{{\ttfamily arXiv:hep-ph/9310320}}.

\bibitem{Dudas:1995yu}
E.~Dudas, S.~Pokorski, and C.~A. Savoy, ``{Yukawa matrices from a spontaneously
  broken Abelian symmetry},''
  \href{http://dx.doi.org/10.1016/0370-2693(95)00795-M}{{\em Phys. Lett. B}
  {\bfseries 356} (1995) 45--55},
  \href{http://arxiv.org/abs/hep-ph/9504292}{{\ttfamily arXiv:hep-ph/9504292}}.

\bibitem{Chen:2012ha}
M.-C. Chen, M.~Fallbacher, M.~Ratz, and C.~Staudt, ``{On predictions from
  spontaneously broken flavor symmetries},''
  \href{http://dx.doi.org/10.1016/j.physletb.2012.10.077}{{\em Phys. Lett. B}
  {\bfseries 718} (2012) 516--521},
  \href{http://arxiv.org/abs/1208.2947}{{\ttfamily arXiv:1208.2947 [hep-ph]}}.

\bibitem{Chen:2013aya}
M.-C. Chen, M.~Fallbacher, Y.~Omura, M.~Ratz, and C.~Staudt, ``{Predictivity of
  models with spontaneously broken non-Abelian discrete flavor symmetries},''
  \href{http://dx.doi.org/10.1016/j.nuclphysb.2013.04.020}{{\em Nucl. Phys. B}
  {\bfseries 873} (2013) 343--371},
  \href{http://arxiv.org/abs/1302.5576}{{\ttfamily arXiv:1302.5576 [hep-ph]}}.

\bibitem{Bazzocchi:2007na}
F.~Bazzocchi, S.~Kaneko, and S.~Morisi, ``{A SUSY A(4) model for fermion masses
  and mixings},'' \href{http://dx.doi.org/10.1088/1126-6708/2008/03/063}{{\em
  JHEP} {\bfseries 03} (2008) 063},
  \href{http://arxiv.org/abs/0707.3032}{{\ttfamily arXiv:0707.3032 [hep-ph]}}.

\bibitem{Feruglio:2009iu}
F.~Feruglio, C.~Hagedorn, and L.~Merlo, ``{Vacuum Alignment in SUSY A4
  Models},'' \href{http://dx.doi.org/10.1007/JHEP03(2010)084}{{\em JHEP}
  {\bfseries 03} (2010) 084}, \href{http://arxiv.org/abs/0910.4058}{{\ttfamily
  arXiv:0910.4058 [hep-ph]}}.

\bibitem{King:2011zj}
S.~F. King and C.~Luhn, ``{Trimaximal neutrino mixing from vacuum alignment in
  A4 and S4 models},'' \href{http://dx.doi.org/10.1007/JHEP09(2011)042}{{\em
  JHEP} {\bfseries 09} (2011) 042},
  \href{http://arxiv.org/abs/1107.5332}{{\ttfamily arXiv:1107.5332 [hep-ph]}}.

\bibitem{Holthausen:2011vd}
M.~Holthausen and M.~A. Schmidt, ``{Natural Vacuum Alignment from Group Theory:
  The Minimal Case},'' \href{http://dx.doi.org/10.1007/JHEP01(2012)126}{{\em
  JHEP} {\bfseries 01} (2012) 126},
  \href{http://arxiv.org/abs/1111.1730}{{\ttfamily arXiv:1111.1730 [hep-ph]}}.

\bibitem{Kobayashi:2008ih}
T.~Kobayashi, Y.~Omura, and K.~Yoshioka, ``{Flavor Symmetry Breaking and Vacuum
  Alignment on Orbifolds},''
  \href{http://dx.doi.org/10.1103/PhysRevD.78.115006}{{\em Phys. Rev. D}
  {\bfseries 78} (2008) 115006},
  \href{http://arxiv.org/abs/0809.3064}{{\ttfamily arXiv:0809.3064 [hep-ph]}}.

\bibitem{Chen:2009gf}
M.-C. Chen and K.~T. Mahanthappa, ``{Group Theoretical Origin of CP
  Violation},'' \href{http://dx.doi.org/10.1016/j.physletb.2009.10.059}{{\em
  Phys. Lett. B} {\bfseries 681} (2009) 444--447},
  \href{http://arxiv.org/abs/0904.1721}{{\ttfamily arXiv:0904.1721 [hep-ph]}}.

\bibitem{Chen:2014tpa}
M.-C. Chen, M.~Fallbacher, K.~T. Mahanthappa, M.~Ratz, and A.~Trautner, ``{CP
  Violation from Finite Groups},''
  \href{http://dx.doi.org/10.1016/j.nuclphysb.2014.03.023}{{\em Nucl. Phys. B}
  {\bfseries 883} (2014) 267--305},
  \href{http://arxiv.org/abs/1402.0507}{{\ttfamily arXiv:1402.0507 [hep-ph]}}.

\bibitem{Feruglio:2012cw}
F.~Feruglio, C.~Hagedorn, and R.~Ziegler, ``{Lepton Mixing Parameters from
  Discrete and CP Symmetries},''
  \href{http://dx.doi.org/10.1007/JHEP07(2013)027}{{\em JHEP} {\bfseries 07}
  (2013) 027}, \href{http://arxiv.org/abs/1211.5560}{{\ttfamily arXiv:1211.5560
  [hep-ph]}}.

\bibitem{Holthausen:2012dk}
M.~Holthausen, M.~Lindner, and M.~A. Schmidt, ``{CP and Discrete Flavour
  Symmetries},'' \href{http://dx.doi.org/10.1007/JHEP04(2013)122}{{\em JHEP}
  {\bfseries 04} (2013) 122}, \href{http://arxiv.org/abs/1211.6953}{{\ttfamily
  arXiv:1211.6953 [hep-ph]}}.

\bibitem{Trautner:2016ezn}
A.~Trautner, {\em {CP and other Symmetries of Symmetries}}.
\newblock PhD thesis, Munich, Tech. U., Universe, 2016.
\newblock \href{http://arxiv.org/abs/1608.05240}{{\ttfamily arXiv:1608.05240
  [hep-ph]}}.
\newblock {\texttt{https://arxiv.org/abs/1608.05240}}.

\bibitem{Kobayashi:2006wq}
T.~Kobayashi, H.~P. Nilles, F.~Pl{\"o}ger, S.~Raby, and M.~Ratz, ``{Stringy
  origin of non-Abelian discrete flavor symmetries},''
  \href{http://dx.doi.org/10.1016/j.nuclphysb.2007.01.018}{{\em Nucl. Phys. B}
  {\bfseries 768} (2007) 135--156},
  \href{http://arxiv.org/abs/hep-ph/0611020}{{\ttfamily arXiv:hep-ph/0611020}}.

\bibitem{Ibanez:1987sn}
L.~E. Ib\'{a}\~{n}ez, J.~E. Kim, H.~P. Nilles, and F.~Quevedo, ``{Orbifold
  Compactifications with Three Families of SU(3) x SU(2) x U(1)**n},''
  \href{http://dx.doi.org/10.1016/0370-2693(87)90255-3}{{\em Phys. Lett. B}
  {\bfseries 191} (1987) 282--286}.

\bibitem{Nilles:2018wex}
H.~P. Nilles, M.~Ratz, A.~Trautner, and P.~K.~S. Vaudrevange, ``{$\mathcal{CP}$
  violation from string theory},''
  \href{http://dx.doi.org/10.1016/j.physletb.2018.09.053}{{\em Phys. Lett. B}
  {\bfseries 786} (2018) 283--287},
  \href{http://arxiv.org/abs/1808.07060}{{\ttfamily arXiv:1808.07060
  [hep-th]}}.

\bibitem{Ratz:2016scn}
M.~Ratz and A.~Trautner, ``{$\mathcal{CP}$ violation with an unbroken
  $\mathcal{CP}$ transformation},''
  \href{http://dx.doi.org/10.1007/JHEP02(2017)103}{{\em JHEP} {\bfseries 02}
  (2017) 103}, \href{http://arxiv.org/abs/1612.08984}{{\ttfamily
  arXiv:1612.08984 [hep-ph]}}.

\bibitem{Abe:2009vi}
H.~Abe, K.-S. Choi, T.~Kobayashi, and H.~Ohki, ``{Non-Abelian Discrete Flavor
  Symmetries from Magnetized/Intersecting Brane Models},''
  \href{http://dx.doi.org/10.1016/j.nuclphysb.2009.05.024}{{\em Nucl. Phys. B}
  {\bfseries 820} (2009) 317--333},
  \href{http://arxiv.org/abs/0904.2631}{{\ttfamily arXiv:0904.2631 [hep-ph]}}.

\bibitem{Cvetic:2018xaq}
M.~Cveti\v{c}, J.~J. Heckman, and L.~Lin, ``{Towards Exotic Matter and Discrete
  Non-Abelian Symmetries in F-theory},''
  \href{http://dx.doi.org/10.1007/JHEP11(2018)001}{{\em JHEP} {\bfseries 11}
  (2018) 001}, \href{http://arxiv.org/abs/1806.10594}{{\ttfamily
  arXiv:1806.10594 [hep-th]}}.

\bibitem{Kobayashi:2018vbk}
T.~Kobayashi, K.~Tanaka, and T.~H. Tatsuishi, ``{Neutrino mixing from finite
  modular groups},'' \href{http://dx.doi.org/10.1103/PhysRevD.98.016004}{{\em
  Phys. Rev. D} {\bfseries 98} no.~1, (2018) 016004},
  \href{http://arxiv.org/abs/1803.10391}{{\ttfamily arXiv:1803.10391
  [hep-ph]}}.

\bibitem{deAnda:2018ecu}
F.~J. de~Anda, S.~F. King, and E.~Perdomo, ``{$SU(5)$ grand unified theory with
  $A_4$ modular symmetry},''
  \href{http://dx.doi.org/10.1103/PhysRevD.101.015028}{{\em Phys. Rev. D}
  {\bfseries 101} no.~1, (2020) 015028},
  \href{http://arxiv.org/abs/1812.05620}{{\ttfamily arXiv:1812.05620
  [hep-ph]}}.

\bibitem{Okada:2018yrn}
H.~Okada and M.~Tanimoto, ``{CP violation of quarks in $A_4$ modular
  invariance},'' \href{http://dx.doi.org/10.1016/j.physletb.2019.02.028}{{\em
  Phys. Lett. B} {\bfseries 791} (2019) 54--61},
  \href{http://arxiv.org/abs/1812.09677}{{\ttfamily arXiv:1812.09677
  [hep-ph]}}.

\bibitem{Ding:2019xna}
G.-J. Ding, S.~F. King, and X.-G. Liu, ``{Neutrino mass and mixing with $A_5$
  modular symmetry},''
  \href{http://dx.doi.org/10.1103/PhysRevD.100.115005}{{\em Phys. Rev. D}
  {\bfseries 100} no.~11, (2019) 115005},
  \href{http://arxiv.org/abs/1903.12588}{{\ttfamily arXiv:1903.12588
  [hep-ph]}}.

\bibitem{Novichkov:2019sqv}
P.~Novichkov, J.~Penedo, S.~Petcov, and A.~Titov, ``{Generalised CP Symmetry in
  Modular-Invariant Models of Flavour},''
  \href{http://dx.doi.org/10.1007/JHEP07(2019)165}{{\em JHEP} {\bfseries 07}
  (2019) 165}, \href{http://arxiv.org/abs/1905.11970}{{\ttfamily
  arXiv:1905.11970 [hep-ph]}}.

\bibitem{Liu:2019khw}
X.-G. Liu and G.-J. Ding, ``{Neutrino Masses and Mixing from Double Covering of
  Finite Modular Groups},''
  \href{http://dx.doi.org/10.1007/JHEP08(2019)134}{{\em JHEP} {\bfseries 08}
  (2019) 134}, \href{http://arxiv.org/abs/1907.01488}{{\ttfamily
  arXiv:1907.01488 [hep-ph]}}.

\bibitem{Kobayashi:2019xvz}
T.~Kobayashi, Y.~Shimizu, K.~Takagi, M.~Tanimoto, and T.~H. Tatsuishi, ``{$A_4$
  lepton flavor model and modulus stabilization from $S_4$ modular symmetry},''
  \href{http://dx.doi.org/10.1103/PhysRevD.100.115045}{{\em Phys. Rev. D}
  {\bfseries 100} no.~11, (2019) 115045},
  \href{http://arxiv.org/abs/1909.05139}{{\ttfamily arXiv:1909.05139
  [hep-ph]}}. [Erratum: Phys.Rev.D 101, 039904 (2020)].

\bibitem{Asaka:2019vev}
T.~Asaka, Y.~Heo, T.~H. Tatsuishi, and T.~Yoshida, ``{Modular $A_4$ invariance
  and leptogenesis},'' \href{http://dx.doi.org/10.1007/JHEP01(2020)144}{{\em
  JHEP} {\bfseries 01} (2020) 144},
  \href{http://arxiv.org/abs/1909.06520}{{\ttfamily arXiv:1909.06520
  [hep-ph]}}.

\bibitem{Gui-JunDing:2019wap}
G.-J. Ding, S.~F. King, X.-G. Liu, and J.-N. Lu, ``{Modular S$_{4}$ and A$_{4}$
  symmetries and their fixed points: new predictive examples of lepton
  mixing},'' \href{http://dx.doi.org/10.1007/JHEP12(2019)030}{{\em JHEP}
  {\bfseries 12} (2019) 030}, \href{http://arxiv.org/abs/1910.03460}{{\ttfamily
  arXiv:1910.03460 [hep-ph]}}.

\bibitem{Kobayashi:2019uyt}
T.~Kobayashi, Y.~Shimizu, K.~Takagi, M.~Tanimoto, T.~H. Tatsuishi, and
  H.~Uchida, ``{$CP$ violation in modular invariant flavor models},''
  \href{http://dx.doi.org/10.1103/PhysRevD.101.055046}{{\em Phys. Rev. D}
  {\bfseries 101} no.~5, (2020) 055046},
  \href{http://arxiv.org/abs/1910.11553}{{\ttfamily arXiv:1910.11553
  [hep-ph]}}.

\bibitem{Ding:2020yen}
G.-J. Ding and F.~Feruglio, ``{Testing Moduli and Flavon Dynamics with Neutrino
  Oscillations},'' \href{http://dx.doi.org/10.1007/JHEP06(2020)134}{{\em JHEP}
  {\bfseries 06} (2020) 134}, \href{http://arxiv.org/abs/2003.13448}{{\ttfamily
  arXiv:2003.13448 [hep-ph]}}.

\bibitem{Liu:2020msy}
X.-G. Liu, C.-Y. Yao, B.-Y. Qu, and G.-J. Ding, ``{Half-integral weight modular
  forms and application to neutrino mass models},''
  \href{http://dx.doi.org/10.1103/PhysRevD.102.115035}{{\em Phys. Rev. D}
  {\bfseries 102} no.~11, (2020) 115035},
  \href{http://arxiv.org/abs/2007.13706}{{\ttfamily arXiv:2007.13706
  [hep-ph]}}.

\bibitem{Ding:2020zxw}
G.-J. Ding, F.~Feruglio, and X.-G. Liu, ``{Automorphic Forms and Fermion
  Masses},'' \href{http://dx.doi.org/10.1007/JHEP01(2021)037}{{\em JHEP}
  {\bfseries 01} (2021) 037}, \href{http://arxiv.org/abs/2010.07952}{{\ttfamily
  arXiv:2010.07952 [hep-th]}}.

\bibitem{Yao:2020zml}
C.-Y. Yao, X.-G. Liu, and G.-J. Ding, ``{Fermion masses and mixing from the
  double cover and metaplectic cover of the $A_5$ modular group},''
  \href{http://dx.doi.org/10.1103/PhysRevD.103.095013}{{\em Phys. Rev. D}
  {\bfseries 103} no.~9, (2021) 095013},
  \href{http://arxiv.org/abs/2011.03501}{{\ttfamily arXiv:2011.03501
  [hep-ph]}}.

\bibitem{Novichkov:2021cgl}
P.~Novichkov, {\em {Aspects of the Modular Symmetry Approach to Lepton
  Flavour}}.
\newblock PhD thesis, SISSA, Trieste, 2021.

\bibitem{Kikuchi:2021ogn}
S.~Kikuchi, T.~Kobayashi, and H.~Uchida, ``{Modular flavor symmetries of
  three-generation modes on magnetized toroidal orbifolds},''
  \href{http://dx.doi.org/10.1103/PhysRevD.104.065008}{{\em Phys. Rev. D}
  {\bfseries 104} no.~6, (2021) 065008},
  \href{http://arxiv.org/abs/2101.00826}{{\ttfamily arXiv:2101.00826
  [hep-th]}}.

\bibitem{Liu:2021gwa}
X.-G. Liu and G.-J. Ding, ``{Modular flavor symmetry and vector-valued modular
  forms},'' \href{http://dx.doi.org/10.1007/JHEP03(2022)123}{{\em JHEP}
  {\bfseries 03} (2022) 123}, \href{http://arxiv.org/abs/2112.14761}{{\ttfamily
  arXiv:2112.14761 [hep-ph]}}.

\bibitem{Dine:1986zy}
M.~Dine, N.~Seiberg, X.~G. Wen, and E.~Witten, ``{Nonperturbative Effects on
  the String World Sheet},''
  \href{http://dx.doi.org/10.1016/0550-3213(86)90418-9}{{\em Nucl. Phys. B}
  {\bfseries 278} (1986) 769--789}.

\bibitem{Dine:1987bq}
M.~Dine, N.~Seiberg, X.~G. Wen, and E.~Witten, ``{Nonperturbative Effects on
  the String World Sheet. 2.},''
  \href{http://dx.doi.org/10.1016/0550-3213(87)90383-X}{{\em Nucl. Phys. B}
  {\bfseries 289} (1987) 319--363}.

\bibitem{Cvetic:1987qx}
M.~Cveti{\v{c}}, ``{Suppression of Nonrenormalizable Terms in the Effective
  Superpotential for (Blownup) Orbifold Compactification},''
  \href{http://dx.doi.org/10.1103/PhysRevLett.59.1795}{{\em Phys. Rev. Lett.}
  {\bfseries 59} (1987) 1795}.

\bibitem{Font:1990nt}
A.~Font, L.~E. Ib{\'a}{\~n}ez, D.~L{\"u}st, and F.~Quevedo, ``{Supersymmetry
  Breaking From Duality Invariant Gaugino Condensation},''
  \href{http://dx.doi.org/10.1016/0370-2693(90)90665-S}{{\em Phys. Lett. B}
  {\bfseries 245} (1990) 401--408}.

\bibitem{Nilles:1990jv}
H.~P. Nilles and M.~Olechowski, ``{Gaugino Condensation and Duality
  Invariance},'' \href{http://dx.doi.org/10.1016/0370-2693(90)90290-M}{{\em
  Phys. Lett. B} {\bfseries 248} (1990) 268--272}.

\bibitem{Novichkov:2022wvg}
P.~P. Novichkov, J.~T. Penedo, and S.~T. Petcov, ``{Modular Flavour Symmetries
  and Modulus Stabilisation},''
  \href{http://arxiv.org/abs/2201.02020}{{\ttfamily arXiv:2201.02020
  [hep-ph]}}.

\bibitem{Okada:2020ukr}
H.~Okada and M.~Tanimoto, ``{Modular invariant flavor model of $A_4$ and
  hierarchical structures at nearby fixed points},''
  \href{http://dx.doi.org/10.1103/PhysRevD.103.015005}{{\em Phys. Rev. D}
  {\bfseries 103} no.~1, (2021) 015005},
  \href{http://arxiv.org/abs/2009.14242}{{\ttfamily arXiv:2009.14242
  [hep-ph]}}.

\bibitem{Feruglio:2021dte}
F.~Feruglio, V.~Gherardi, A.~Romanino, and A.~Titov, ``{Modular invariant
  dynamics and fermion mass hierarchies around $\tau = i$},''
  \href{http://dx.doi.org/10.1007/JHEP05(2021)242}{{\em JHEP} {\bfseries 05}
  (2021) 242}, \href{http://arxiv.org/abs/2101.08718}{{\ttfamily
  arXiv:2101.08718 [hep-ph]}}.

\bibitem{Novichkov:2021evw}
P.~P. Novichkov, J.~T. Penedo, and S.~T. Petcov, ``{Fermion mass hierarchies,
  large lepton mixing and residual modular symmetries},''
  \href{http://dx.doi.org/10.1007/JHEP04(2021)206}{{\em JHEP} {\bfseries 04}
  (2021) 206}, \href{http://arxiv.org/abs/2102.07488}{{\ttfamily
  arXiv:2102.07488 [hep-ph]}}.

\bibitem{Chen:2019ewa}
M.-C. Chen, S.~Ramos-S{\'a}nchez, and M.~Ratz, ``{A note on the predictions of
  models with modular flavor symmetries},''
  \href{http://dx.doi.org/10.1016/j.physletb.2019.135153}{{\em Phys. Lett. B}
  {\bfseries 801} (2020) 135153},
  \href{http://arxiv.org/abs/1909.06910}{{\ttfamily arXiv:1909.06910
  [hep-ph]}}.

\bibitem{Chen:2021prl}
M.-C. Chen, V.~Knapp-P{\'e}rez, M.~Ramos-Hamud, S.~Ramos-S{\'a}nchez, M.~Ratz,
  and S.~Shukla, ``{Quasi-eclectic modular flavor symmetries},''
  \href{http://dx.doi.org/10.1016/j.physletb.2021.136843}{{\em Phys. Lett. B}
  {\bfseries 824} (2022) 136843},
  \href{http://arxiv.org/abs/2108.02240}{{\ttfamily arXiv:2108.02240
  [hep-ph]}}.

\bibitem{Cremades:2004wa}
D.~Cremades, L.~E. Ib{\'a}{\~n}ez, and F.~Marchesano, ``{Computing Yukawa
  couplings from magnetized extra dimensions},''
  \href{http://dx.doi.org/10.1088/1126-6708/2004/05/079}{{\em JHEP} {\bfseries
  05} (2004) 079}, \href{http://arxiv.org/abs/hep-th/0404229}{{\ttfamily
  arXiv:hep-th/0404229}}.

\bibitem{Cremades:2003qj}
D.~Cremades, L.~E. Ib{\'a}{\~n}ez, and F.~Marchesano, ``{Yukawa couplings in
  intersecting D-brane models},''
  \href{http://dx.doi.org/10.1088/1126-6708/2003/07/038}{{\em JHEP} {\bfseries
  07} (2003) 038}, \href{http://arxiv.org/abs/hep-th/0302105}{{\ttfamily
  arXiv:hep-th/0302105}}.

\bibitem{Kobayashi:2016ovu}
T.~Kobayashi, S.~Nagamoto, and S.~Uemura, ``{Modular symmetry in
  magnetized/intersecting D-brane models},''
  \href{http://dx.doi.org/10.1093/ptep/ptw184}{{\em PTEP} {\bfseries 2017}
  no.~2, (2017) 023B02}, \href{http://arxiv.org/abs/1608.06129}{{\ttfamily
  arXiv:1608.06129 [hep-th]}}.

\bibitem{Kobayashi:2018rad}
T.~Kobayashi, S.~Nagamoto, S.~Takada, S.~Tamba, and T.~H. Tatsuishi, ``{Modular
  symmetry and non-Abelian discrete flavor symmetries in string
  compactification},'' \href{http://dx.doi.org/10.1103/PhysRevD.97.116002}{{\em
  Phys. Rev. D} {\bfseries 97} no.~11, (2018) 116002},
  \href{http://arxiv.org/abs/1804.06644}{{\ttfamily arXiv:1804.06644
  [hep-th]}}.

\bibitem{Kobayashi:2018bff}
T.~Kobayashi and S.~Tamba, ``{Modular forms of finite modular subgroups from
  magnetized D-brane models},''
  \href{http://dx.doi.org/10.1103/PhysRevD.99.046001}{{\em Phys. Rev. D}
  {\bfseries 99} no.~4, (2019) 046001},
  \href{http://arxiv.org/abs/1811.11384}{{\ttfamily arXiv:1811.11384
  [hep-th]}}.

\bibitem{Kariyazono:2019ehj}
Y.~Kariyazono, T.~Kobayashi, S.~Takada, S.~Tamba, and H.~Uchida, ``{Modular
  symmetry anomaly in magnetic flux compactification},''
  \href{http://dx.doi.org/10.1103/PhysRevD.100.045014}{{\em Phys. Rev. D}
  {\bfseries 100} no.~4, (2019) 045014},
  \href{http://arxiv.org/abs/1904.07546}{{\ttfamily arXiv:1904.07546
  [hep-th]}}.

\bibitem{Ohki:2020bpo}
H.~Ohki, S.~Uemura, and R.~Watanabe, ``{Modular flavor symmetry on a magnetized
  torus},'' \href{http://dx.doi.org/10.1103/PhysRevD.102.085008}{{\em Phys.
  Rev. D} {\bfseries 102} no.~8, (2020) 085008},
  \href{http://arxiv.org/abs/2003.04174}{{\ttfamily arXiv:2003.04174
  [hep-th]}}.

\bibitem{Kikuchi:2020frp}
S.~Kikuchi, T.~Kobayashi, S.~Takada, T.~H. Tatsuishi, and H.~Uchida,
  ``{Revisiting modular symmetry in magnetized torus and orbifold
  compactifications},''
  \href{http://dx.doi.org/10.1103/PhysRevD.102.105010}{{\em Phys. Rev. D}
  {\bfseries 102} no.~10, (2020) 105010},
  \href{http://arxiv.org/abs/2005.12642}{{\ttfamily arXiv:2005.12642
  [hep-th]}}.

\bibitem{Kikuchi:2020nxn}
S.~Kikuchi, T.~Kobayashi, H.~Otsuka, S.~Takada, and H.~Uchida, ``{Modular
  symmetry by orbifolding magnetized $T^2\times T^2$: realization of double
  cover of $\Gamma_N$},'' \href{http://dx.doi.org/10.1007/JHEP11(2020)101}{{\em
  JHEP} {\bfseries 11} (2020) 101},
  \href{http://arxiv.org/abs/2007.06188}{{\ttfamily arXiv:2007.06188
  [hep-th]}}.

\bibitem{Almumin:2021fbk}
Y.~Almumin, M.-C. Chen, V.~Knapp-P\'erez, S.~Ramos-S{\'a}nchez, M.~Ratz, and
  S.~Shukla, ``{Metaplectic Flavor Symmetries from Magnetized Tori},''
  \href{http://dx.doi.org/10.1007/JHEP05(2021)078}{{\em JHEP} {\bfseries 05}
  (2021) 078}, \href{http://arxiv.org/abs/2102.11286}{{\ttfamily
  arXiv:2102.11286 [hep-th]}}.

\bibitem{Tatsuta:2021deu}
Y.~Tatsuta, ``{Modular symmetry and zeros in magnetic compactifications},''
  \href{http://dx.doi.org/10.1007/JHEP10(2021)054}{{\em JHEP} {\bfseries 10}
  (2021) 054}, \href{http://arxiv.org/abs/2104.03855}{{\ttfamily
  arXiv:2104.03855 [hep-th]}}.

\bibitem{Kikuchi:2022bkn}
S.~Kikuchi, T.~Kobayashi, K.~Nasu, H.~Uchida, and S.~Uemura, ``{Modular
  symmetry anomaly and non-perturbative neutrino mass terms in magnetized
  orbifold models},'' \href{http://arxiv.org/abs/2202.05425}{{\ttfamily
  arXiv:2202.05425 [hep-th]}}.

\bibitem{Lauer:1989ax}
J.~Lauer, J.~Mas, and H.~P. Nilles, ``{Duality and the Role of Nonperturbative
  Effects on the World Sheet},''
  \href{http://dx.doi.org/10.1016/0370-2693(89)91190-8}{{\em Phys. Lett. B}
  {\bfseries 226} (1989) 251--256}.

\bibitem{Lauer:1990tm}
J.~Lauer, J.~Mas, and H.~P. Nilles, ``{Twisted sector representations of
  discrete background symmetries for two-dimensional orbifolds},''
  \href{http://dx.doi.org/10.1016/0550-3213(91)90095-F}{{\em Nucl. Phys. B}
  {\bfseries 351} (1991) 353--424}.

\bibitem{Bailin:1999nk}
D.~Bailin and A.~Love, ``{Orbifold compactifications of string theory},''
  \href{http://dx.doi.org/10.1016/S0370-1573(98)00126-4}{{\em Phys. Rept.}
  {\bfseries 315} (1999) 285--408}.

\bibitem{Ramos-Sanchez:2008nwx}
S.~Ramos-S{\'a}nchez, ``{Towards Low Energy Physics from the Heterotic
  String},'' \href{http://dx.doi.org/10.1002/prop.200900073}{{\em Fortsch.
  Phys.} {\bfseries 10} (2009) 907--1036},
  \href{http://arxiv.org/abs/0812.3560}{{\ttfamily arXiv:0812.3560 [hep-th]}}.

\bibitem{Vaudrevange:2008sm}
P.~K.~S. Vaudrevange, {\em {Grand Unification in the Heterotic Brane World}}.
\newblock PhD thesis, Bonn U., 2008.
\newblock \href{http://arxiv.org/abs/0812.3503}{{\ttfamily arXiv:0812.3503
  [hep-th]}}.

\bibitem{Nilles:2008gq}
H.~P. Nilles, S.~Ramos-S{\'a}nchez, M.~Ratz, and P.~K.~S. Vaudrevange, ``{From
  strings to the MSSM},''
  \href{http://dx.doi.org/10.1140/epjc/s10052-008-0740-1}{{\em Eur. Phys. J. C}
  {\bfseries 59} (2009) 249--267},
  \href{http://arxiv.org/abs/0806.3905}{{\ttfamily arXiv:0806.3905 [hep-th]}}.

\bibitem{Kappl:2010yu}
R.~Kappl, B.~Petersen, S.~Raby, M.~Ratz, R.~Schieren, and P.~K.~S. Vaudrevange,
  ``{String-Derived MSSM Vacua with Residual R Symmetries},''
  \href{http://dx.doi.org/10.1016/j.nuclphysb.2011.01.032}{{\em Nucl. Phys. B}
  {\bfseries 847} (2011) 325--349},
  \href{http://arxiv.org/abs/1012.4574}{{\ttfamily arXiv:1012.4574 [hep-th]}}.

\bibitem{Nilles:2014owa}
H.~P. Nilles and P.~K.~S. Vaudrevange, ``{Geography of Fields in Extra
  Dimensions: String Theory Lessons for Particle Physics},''
  \href{http://dx.doi.org/10.1142/S0217732315300086}{{\em Mod. Phys. Lett. A}
  {\bfseries 30} no.~10, (2015) 1530008},
  \href{http://arxiv.org/abs/1403.1597}{{\ttfamily arXiv:1403.1597 [hep-th]}}.

\bibitem{Parr:2019bta}
E.~Parr and P.~K.~S. Vaudrevange, ``{Contrast data mining for the MSSM from
  strings},'' \href{http://dx.doi.org/10.1016/j.nuclphysb.2020.114922}{{\em
  Nucl. Phys. B} {\bfseries 952} (2020) 114922},
  \href{http://arxiv.org/abs/1910.13473}{{\ttfamily arXiv:1910.13473
  [hep-th]}}.

\bibitem{Nilles:2020gvu}
H.~P. Nilles, S.~Ramos-S{\'a}nchez, and P.~K.~S. Vaudrevange, ``{Eclectic
  flavor scheme from ten-dimensional string theory - II detailed technical
  analysis},'' \href{http://dx.doi.org/10.1016/j.nuclphysb.2021.115367}{{\em
  Nucl. Phys. B} {\bfseries 966} (2021) 115367},
  \href{http://arxiv.org/abs/2010.13798}{{\ttfamily arXiv:2010.13798
  [hep-th]}}.

\bibitem{Baur:2020jwc}
A.~Baur, M.~Kade, H.~P. Nilles, S.~Ramos-S{\'a}nchez, and P.~K.~S. Vaudrevange,
  ``{The eclectic flavor symmetry of the $\mathbb{Z}_2$ orbifold},''
  \href{http://dx.doi.org/10.1007/JHEP02(2021)018}{{\em JHEP} {\bfseries 02}
  (2021) 018}, \href{http://arxiv.org/abs/2008.07534}{{\ttfamily
  arXiv:2008.07534 [hep-th]}}.

\bibitem{Nilles:2020tdp}
H.~P. Nilles, S.~Ramos-S{\'a}nchez, and P.~K.~S. Vaudrevange, ``{Eclectic
  flavor scheme from ten-dimensional string theory -- I. Basic results},''
  \href{http://dx.doi.org/10.1016/j.physletb.2020.135615}{{\em Phys. Lett. B}
  {\bfseries 808} (2020) 135615},
  \href{http://arxiv.org/abs/2006.03059}{{\ttfamily arXiv:2006.03059
  [hep-th]}}.

\bibitem{Nilles:2020kgo}
H.~P. Nilles, S.~Ramos-S{\'a}nchez, and P.~K.~S. Vaudrevange, ``{Lessons from
  eclectic flavor symmetries},''
  \href{http://dx.doi.org/10.1016/j.nuclphysb.2020.115098}{{\em Nucl. Phys. B}
  {\bfseries 957} (2020) 115098},
  \href{http://arxiv.org/abs/2004.05200}{{\ttfamily arXiv:2004.05200
  [hep-ph]}}.

\bibitem{Baur:2021mtl}
A.~Baur, M.~Kade, H.~P. Nilles, S.~Ramos-S{\'a}nchez, and P.~K.~S. Vaudrevange,
  ``{Completing the eclectic flavor scheme of the $\mathbb{Z}_2$ orbifold},''
  \href{http://dx.doi.org/10.1007/JHEP06(2021)110}{{\em JHEP} {\bfseries 06}
  (2021) 110}, \href{http://arxiv.org/abs/2104.03981}{{\ttfamily
  arXiv:2104.03981 [hep-th]}}.

\bibitem{Baur:2021bly}
A.~Baur, H.~P. Nilles, S.~Ramos-S{\'a}nchez, A.~Trautner, and P.~K.~S.
  Vaudrevange, ``{Top-down anatomy of flavor symmetry breakdown},''
  \href{http://dx.doi.org/10.1103/PhysRevD.105.055018}{{\em Phys. Rev. D}
  {\bfseries 105} no.~5, (2022) 055018},
  \href{http://arxiv.org/abs/2112.06940}{{\ttfamily arXiv:2112.06940
  [hep-th]}}.

\bibitem{Baur:2022toappear}
A.~Baur, H.~P. Nilles, S.~Ramos-S{\'a}nchez, A.~Trautner, and P.~K.~S.
  Vaudrevange. In preparation.

\bibitem{Nilles:2020nnc}
H.~P. Nilles, S.~Ramos-S{\'a}nchez, and P.~K. Vaudrevange, ``{Eclectic Flavor
  Groups},'' \href{http://dx.doi.org/10.1007/JHEP02(2020)045}{{\em JHEP}
  {\bfseries 02} (2020) 045}, \href{http://arxiv.org/abs/2001.01736}{{\ttfamily
  arXiv:2001.01736 [hep-ph]}}.

\bibitem{Blaszczyk:2014qoa}
M.~Blaszczyk, S.~Groot~Nibbelink, O.~Loukas, and S.~Ramos-S{\'a}nchez,
  ``{Non-supersymmetric heterotic model building},''
  \href{http://dx.doi.org/10.1007/JHEP10(2014)119}{{\em JHEP} {\bfseries 10}
  (2014) 119}, \href{http://arxiv.org/abs/1407.6362}{{\ttfamily arXiv:1407.6362
  [hep-th]}}.

\bibitem{Ashfaque:2015vta}
J.~M. Ashfaque, P.~Athanasopoulos, A.~E. Faraggi, and H.~Sonmez,
  ``{Non-Tachyonic Semi-Realistic Non-Supersymmetric Heterotic String Vacua},''
  \href{http://dx.doi.org/10.1140/epjc/s10052-016-4056-2}{{\em Eur. Phys. J. C}
  {\bfseries 76} no.~4, (2016) 208},
  \href{http://arxiv.org/abs/1506.03114}{{\ttfamily arXiv:1506.03114
  [hep-th]}}.

\bibitem{Abel:2015oxa}
S.~Abel, K.~R. Dienes, and E.~Mavroudi, ``{Towards a nonsupersymmetric string
  phenomenology},'' \href{http://dx.doi.org/10.1103/PhysRevD.91.126014}{{\em
  Phys. Rev. D} {\bfseries 91} no.~12, (2015) 126014},
  \href{http://arxiv.org/abs/1502.03087}{{\ttfamily arXiv:1502.03087
  [hep-th]}}.

\bibitem{Perez-Martinez:2021zjj}
R.~P{\'e}rez-Mart{\'i}nez, S.~Ramos-S{\'a}nchez, and P.~K.~S. Vaudrevange,
  ``{Landscape of promising nonsupersymmetric string models},''
  \href{http://dx.doi.org/10.1103/PhysRevD.104.046026}{{\em Phys. Rev. D}
  {\bfseries 104} no.~4, (2021) 046026},
  \href{http://arxiv.org/abs/2105.03460}{{\ttfamily arXiv:2105.03460
  [hep-th]}}.

\bibitem{GrootNibbelink:2017luf}
S.~Groot~Nibbelink, O.~Loukas, A.~M\"utter, E.~Parr, and P.~K.~S. Vaudrevange,
  ``{Tension Between a Vanishing Cosmological Constant and Non-Supersymmetric
  Heterotic Orbifolds},'' \href{http://dx.doi.org/10.1002/prop.202000044}{{\em
  Fortsch. Phys.} {\bfseries 68} no.~7, (2020) 2000044},
  \href{http://arxiv.org/abs/1710.09237}{{\ttfamily arXiv:1710.09237
  [hep-th]}}.

\end{thebibliography}\endgroup
\bibliographystyle{utphys}

\end{document}